\documentclass[12pt]{article}
\usepackage{amssymb}
\usepackage{amsfonts}
\usepackage{indentfirst}
\usepackage{amsopn}
\usepackage{amsmath}
\usepackage{amsthm}
\usepackage{amscd}
\usepackage{color,graphicx}
\usepackage{epstopdf}

\makeatletter\@addtoreset {equation}{section}\makeatother

\begin{document}

\title{\bf \Large Localized structures on librational and rotational travelling waves in the sine--Gordon equation}

\author{\Large Dmitry E. Pelinovsky and Robert E. White \\
Department of Mathematics, McMaster University, \\
Hamilton, Ontario, Canada, L8S 4K1}

\date{}
\maketitle

% Nonlinear Dynamics, Journal of Nonlinear Science, Journal of Nonlinear Math Physics

\begin{abstract}
We derive exact solutions to the sine--Gordon equation describing localized structures on the background of librational and rotational travelling waves. In the case of librational waves, the exact solution represents a localized spike in space-time coordinates (a rogue wave) which decays to the periodic background algebraically fast. In the case of rotational waves, 
the exact solution represents a kink propagating on the periodic background and decaying algebraically in the transverse direction to its propagation. These solutions model the universal patterns in the dynamics of fluxon condensates in the semi-classical limit. The different dynamics is related to different outcomes of modulational stability of the librational and rotational waves. 
\end{abstract}

\section{Introduction}

This paper is inspired by the series of works \cite{Mil1,Mil2,milluby} on dynamics of the sine-Gordon equation in the semi-classical limit. This physical regime is relevant
for propagation of the magnetic flux along superconducting Josephson junctions \cite{supJJ}.
Other physical applications of the sine--Gordon equation include crystal dislocations, DNA double helix, fermions in the quantum field theory, and
structures in galaxies (see reviews in \cite{Kivshar,Kevrekidis}).

The sine--Gordon equation in the semi-classical limit can be written in the form:
\begin{equation}\label{fluxon}
\epsilon^2 u_{TT} - \epsilon^2 u_{XX} + \sin(u) = 0,
\end{equation}
where the subscripts denote partial derivatives of $u = u(X,T)$ and
the parameter $\epsilon$ is small. By using the initial data
with zero displacement and large velocity, $u(X,0) = 0$ and $\epsilon u_T(X,0) = G(X)$,
the authors of \cite{Mil1,Mil2,milluby} studied the sequence $\{ \epsilon_N \}_{N \in \mathbb{N}}$
with $\epsilon_N \to 0$ as $N \to \infty$,
where $\epsilon_N$ is defined from the $N$-soliton (reflectionless) potential
associated with the $\epsilon$-independent velocity profile $G(X)$. The sequence of solutions  was termed as {\em the fluxon condensate}. The regime
of rotational waves with $\| G \|_{L^{\infty}} > 2$ was studied
in \cite{Mil1}, whereas the regime of librational waves with $\| G \|_{L^{\infty}} < 2$
was studied recently in \cite{milluby},
the classification corresponds to the dynamics of a pendulum with an angle $\theta = \theta(t)$ satisfying 
\begin{equation}\label{pendulum}
\theta''(t) + \sin(\theta(t))= 0.
\end{equation}

It was suggested in \cite{Mil2} that the dynamics of fluxon condensates was different between the rotational
and librational regimes. In both cases, the initial evolution in the semi-classical limit can be modeled by
the travelling wave with slowly varying parameters. Dynamics of librational waves is affected by the gradient catastrophe
and the emergence of a universal pattern of rogue waves (localized spikes in space-time on a distributed background)
\cite{milluby}. Dynamics of rotational waves is accompanied by the emergence of a universal pattern of propagating kinks and antikinks
at the interface between the rotational and librational motion of the fluxon condensate \cite{Mil1}.

\begin{figure}[htb!] 
	\includegraphics[width=0.9\textwidth]{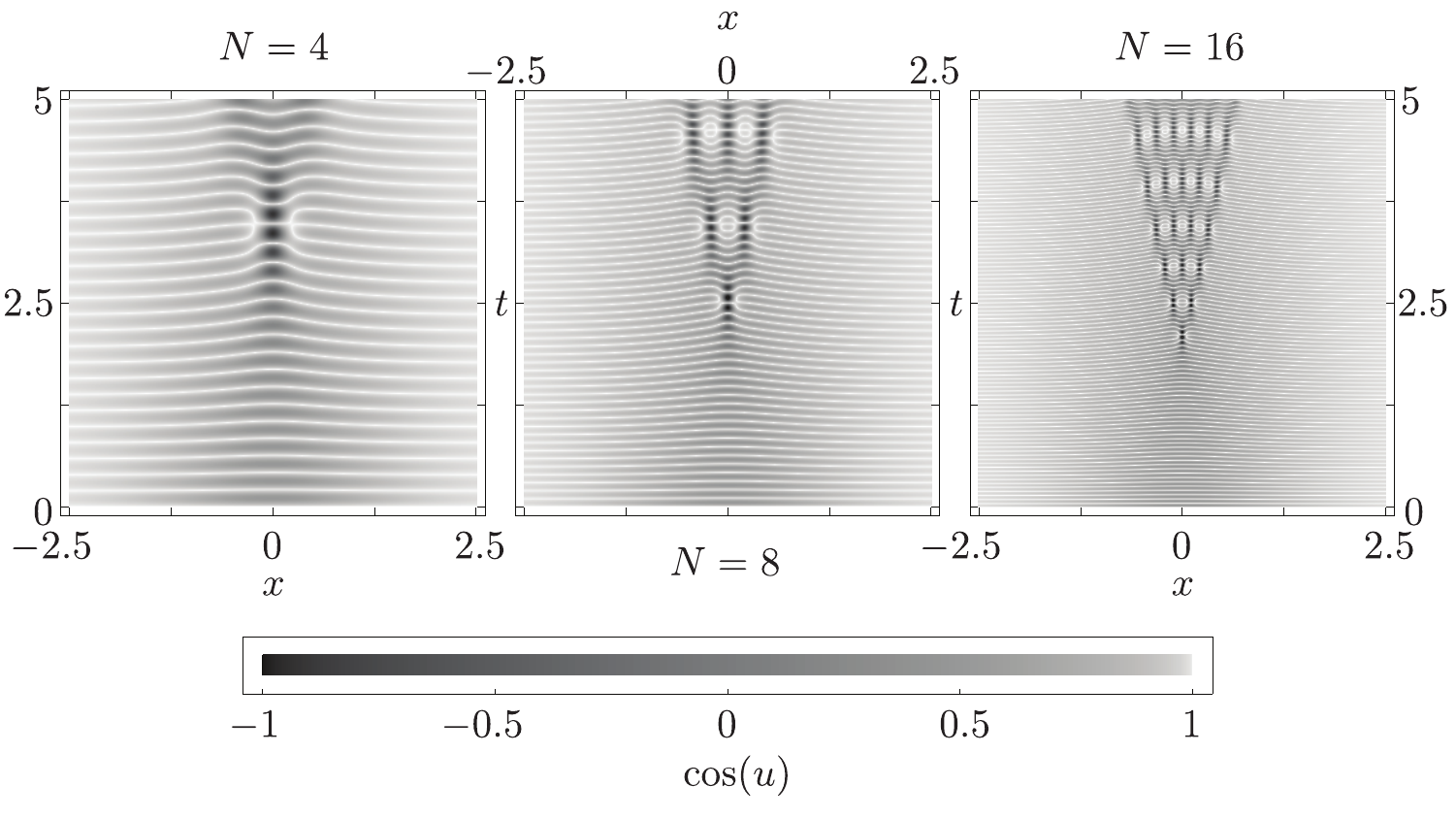}
	\includegraphics[width=0.9\textwidth]{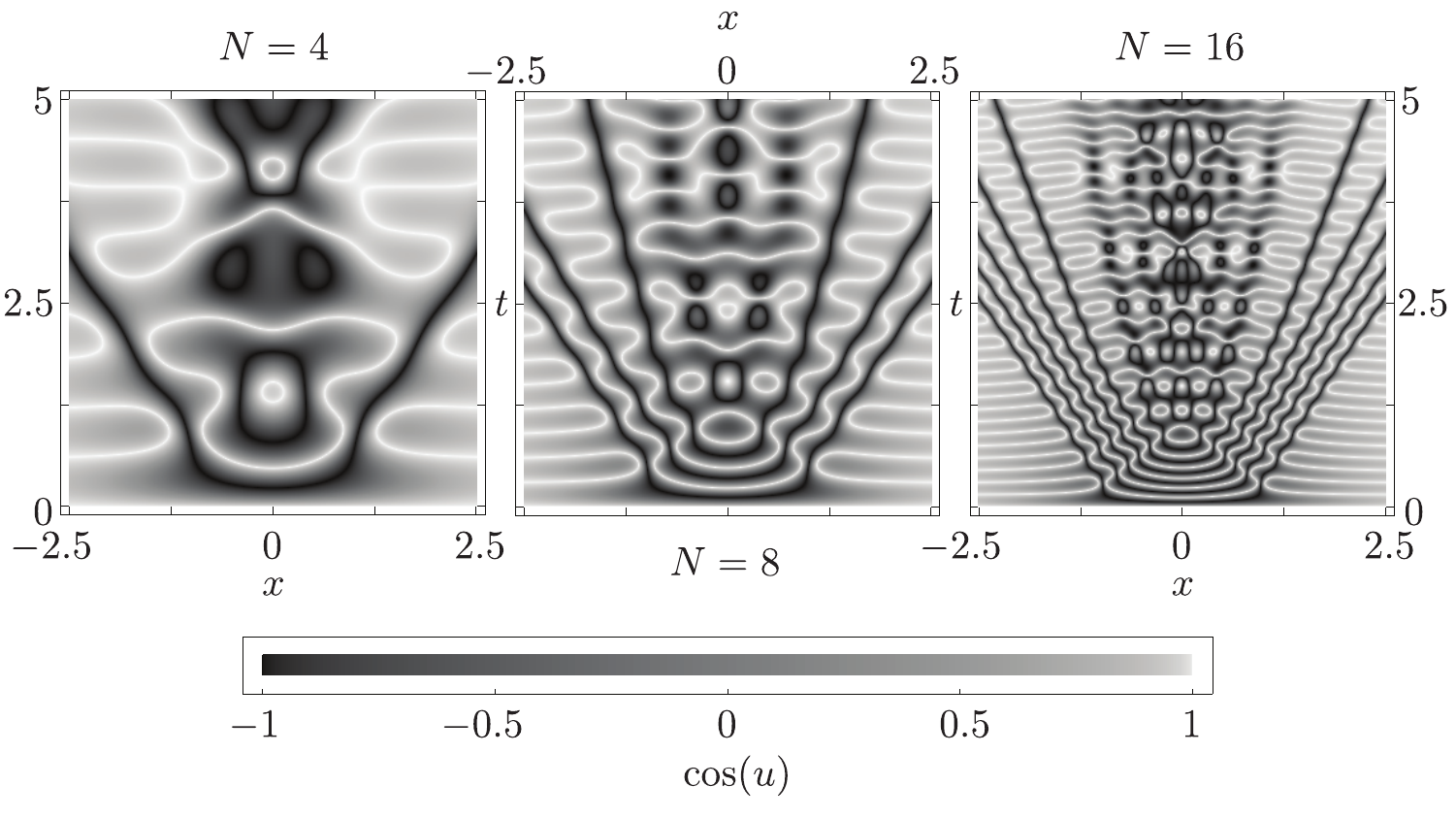}
	\caption{Surface plots of $\cos(u)$ in space-time coordinates 
		for the dynamics of fluxon condensates in the semi-classical limit. 
		Top: librational waves. Bottom: rotational waves. 
		Reproduced from \cite{Mil2} with permission of the authors.}
	\label{fig-miller}
\end{figure}

Figure \ref{fig-miller} (reproduced from \cite{Mil2}) shows the dynamics of $\cos(u)$ in the sine--Gordon equation (\ref{fluxon}) with $\epsilon = \epsilon_N$ for $N = 4, 8, 16$. The top panel shows the regime of librational waves induced by the initial data  $u(X,0) = 0$ and $\epsilon u_T(X,0) = G(X)$ with  $\| G \|_{L^{\infty}} < 2$. The bottom panel shows the regime of rotational waves, for which $\| G \|_{L^{\infty}} > 2$.

The analysis of \cite{Mil1,Mil2,milluby} relies on the reformulation of the Riemann--Hilbert problem used in the integration of the sine--Gordon equation 
and careful asymptotic estimates. The purpose of our work is {\em to develop a short and simple algebraic method},
which allows us to construct the exact solutions for the principal waveforms that make dynamics of librational and rotational waves so different.
In the case of librational waves, we derive a closed-form solution for a rogue wave decaying algebraically
to the periodic background in all directions. In the case of rotational waves, we derive a closed-form solution for propagating kinks and antikinks that decay algebraically to the periodic background in the transverse direction to their propagation.
These solutions with localized structures on librational and rotational waves are associated with the particular eigenvalues in the Lax spectrum for which the eigenfunctions are bounded and periodic in space-time coordinates. Since we are not dealing with the initial-value 
problem in the semi-classical limit, we can scale the space-time coordinates and consider the normalized sine--Gordon equation:
\begin{equation}\label{SG}
u_{tt} - u_{xx} + \sin(u) = 0,
\end{equation}
where $u = u(x,t)$.

Although the algebraic method used for librational and rotational waves is similar, the outcomes are different dynamically.
This difference is explained by the different types of spectral stability of the travelling periodic waves \cite{Jones,Jones2,marangell} 
(see also \cite{DMS,DU2} and \cite{marangell2} for recent contributions).
In the superluminal regime (which is the only regime we are interested in), the librational periodic waves are spectrally unstable and
the Floquet-Bloch spectrum forms a figure eight intersecting at the origin. Such instability is usually referred to as modulational 
instability \cite{Jones,Jones2}. On the other hand, the rotational periodic waves are modulationally stable in the sense that the only Floquet--Bloch 
spectrum near the origin is represented by the vertical bands along the purely imaginary axis. The rotational waves are still spectrally unstable 
in the superluminal regime but the unstable band is given by bubbles away from the origin (see Fig.2 in \cite{Jones}, Figs.1-2 in \cite{marangell}, Fig. 6 in \cite{DMS}, or Fig. 1 in \cite{marangell2}).

We develop the algebraic method which was previously applied to the modified KdV equation in \cite{CPkdv,CPgardner} 
and the focusing cubic NLS equation in \cite{CPnls,CPW1,CPW2}. The travelling periodic waves and the periodic eigenfunctions in space-time coordinates  are characterized by using nonlinearization of the Lax equations \cite{Cao}. This method allows us to find particular eigenvalues in the Lax spectrum, for which the first solutions to the Lax equations are bounded and periodic whereas the second, linearly independent solutions are unbounded and non-periodic. When the second solutions of the Lax equations are used in the Darboux transformation, new solutions of integrable equations are generated from the travelling periodic wave solutions. The new solutions 
represent algebraically localized structures on the background of travelling periodic waves. 
Similar solutions but in a different functional-analytic form were obtained in \cite{Feng} for the NLS equation and in \cite{LiGeng} and \cite{milluby} 
for the sine--Gordon equation.

The algebraic method can be applied similarly to what was done in \cite{CPkdv,CPgardner} because the sine--Gordon equation 
is related to the same Lax spectral problem 
as the modified KdV, the cubic NLS, and other 
integrable equations considered in the seminal work \cite{AKNS}.
In order to enable this application, we have to rewrite the sine--Gordon equation 
in the characteristic coordinates and use the chain rule for the inverse transformation 
of variables. Since many computational details are similar, we will omit many computations 
and refer to \cite{CPkdv,CPgardner} or to \cite{Robert-thesis} where computational details can be found.

\begin{figure}[htb!] 
\includegraphics[width=0.48\textwidth]{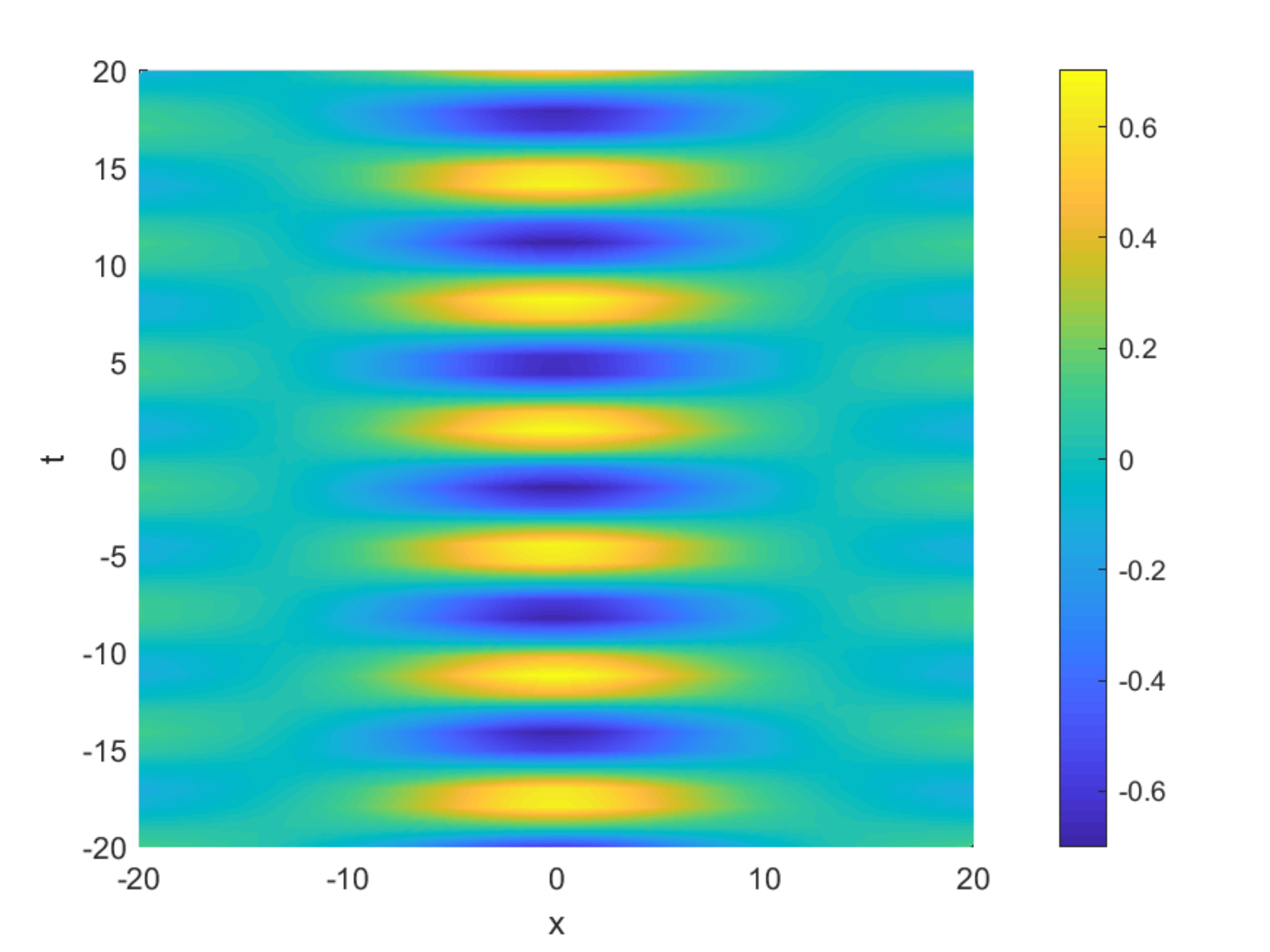}
\includegraphics[width=0.48\textwidth]{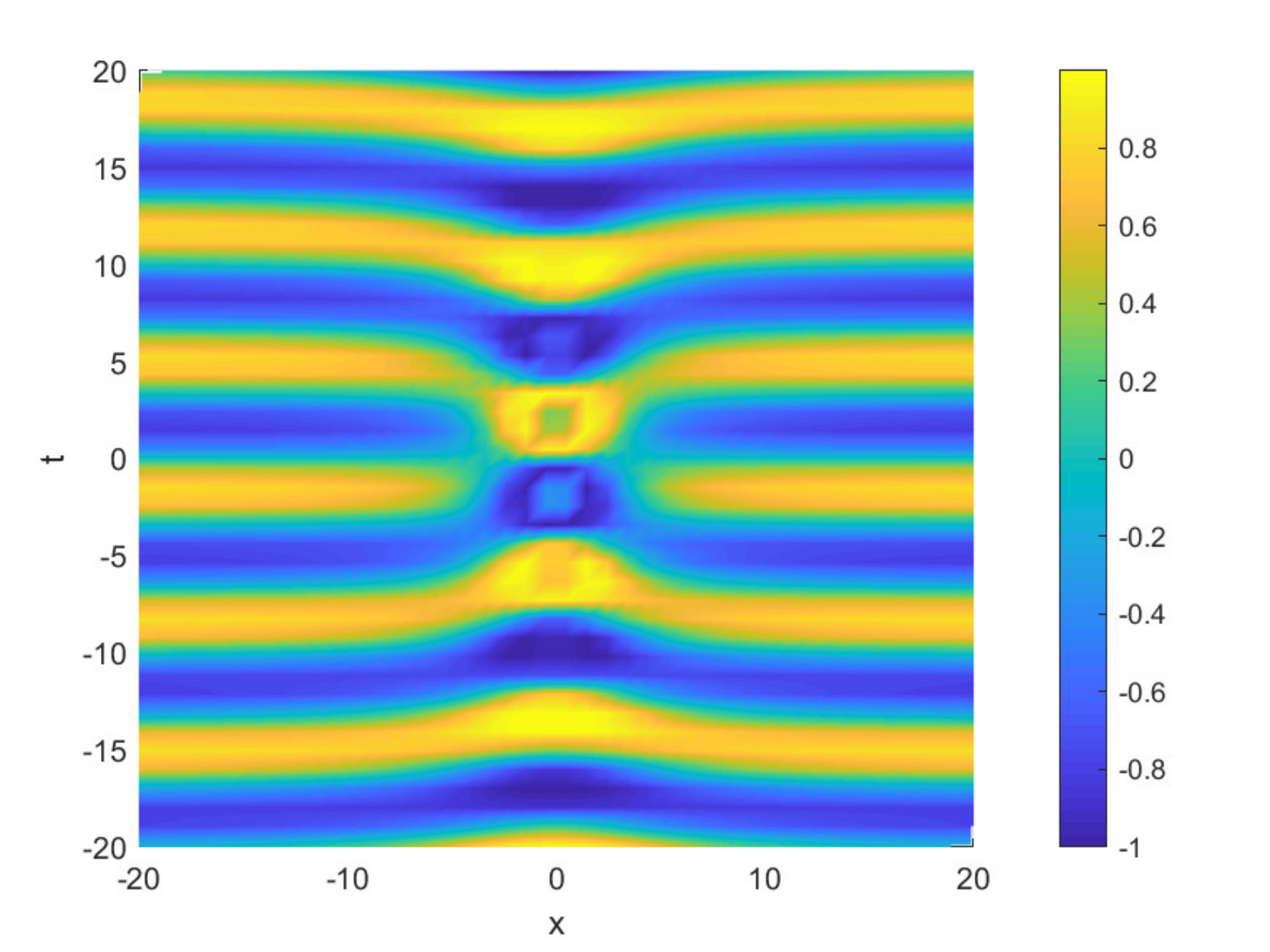}
\includegraphics[width=0.48\textwidth]{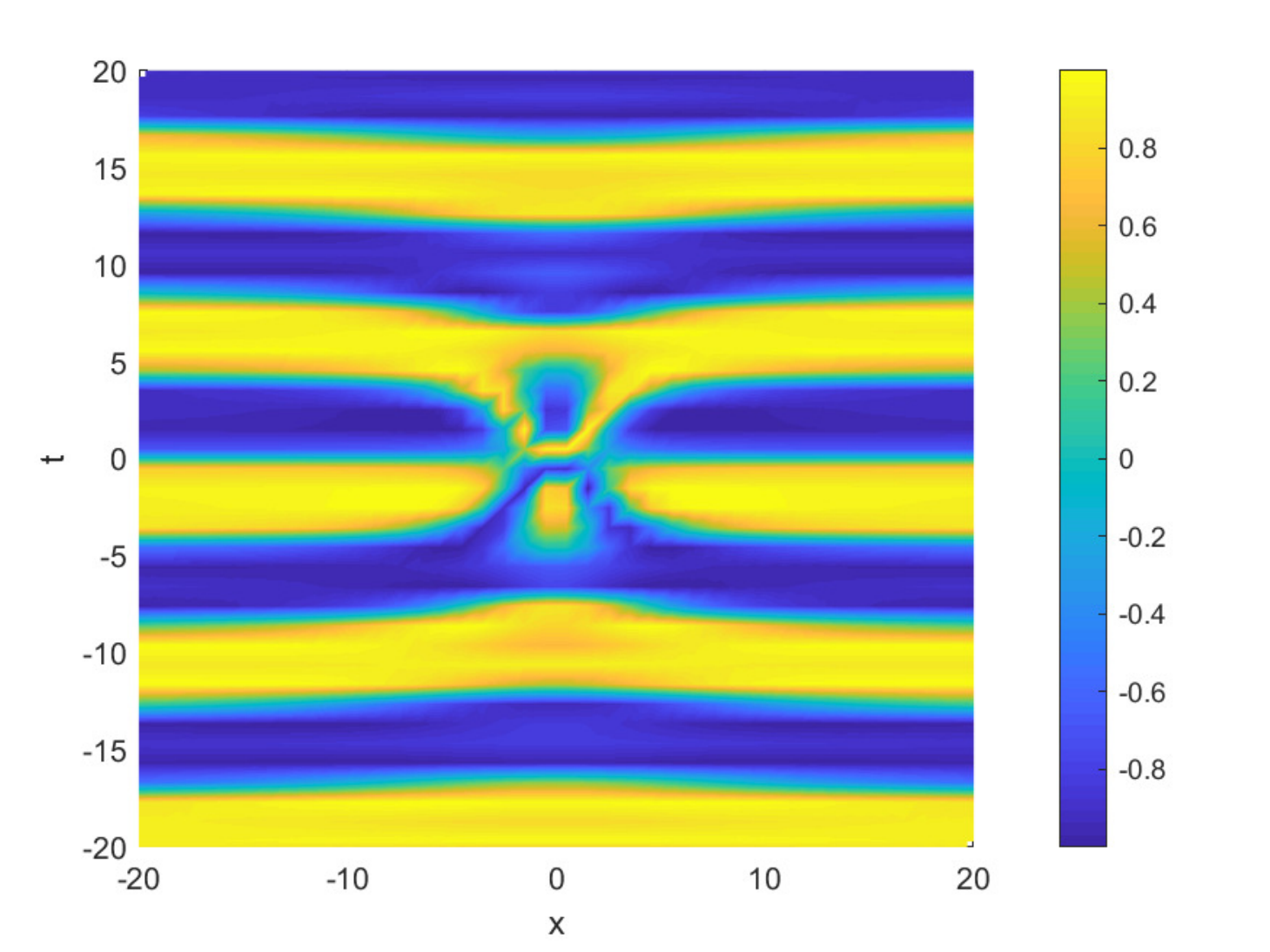} 
\includegraphics[width=0.48\textwidth]{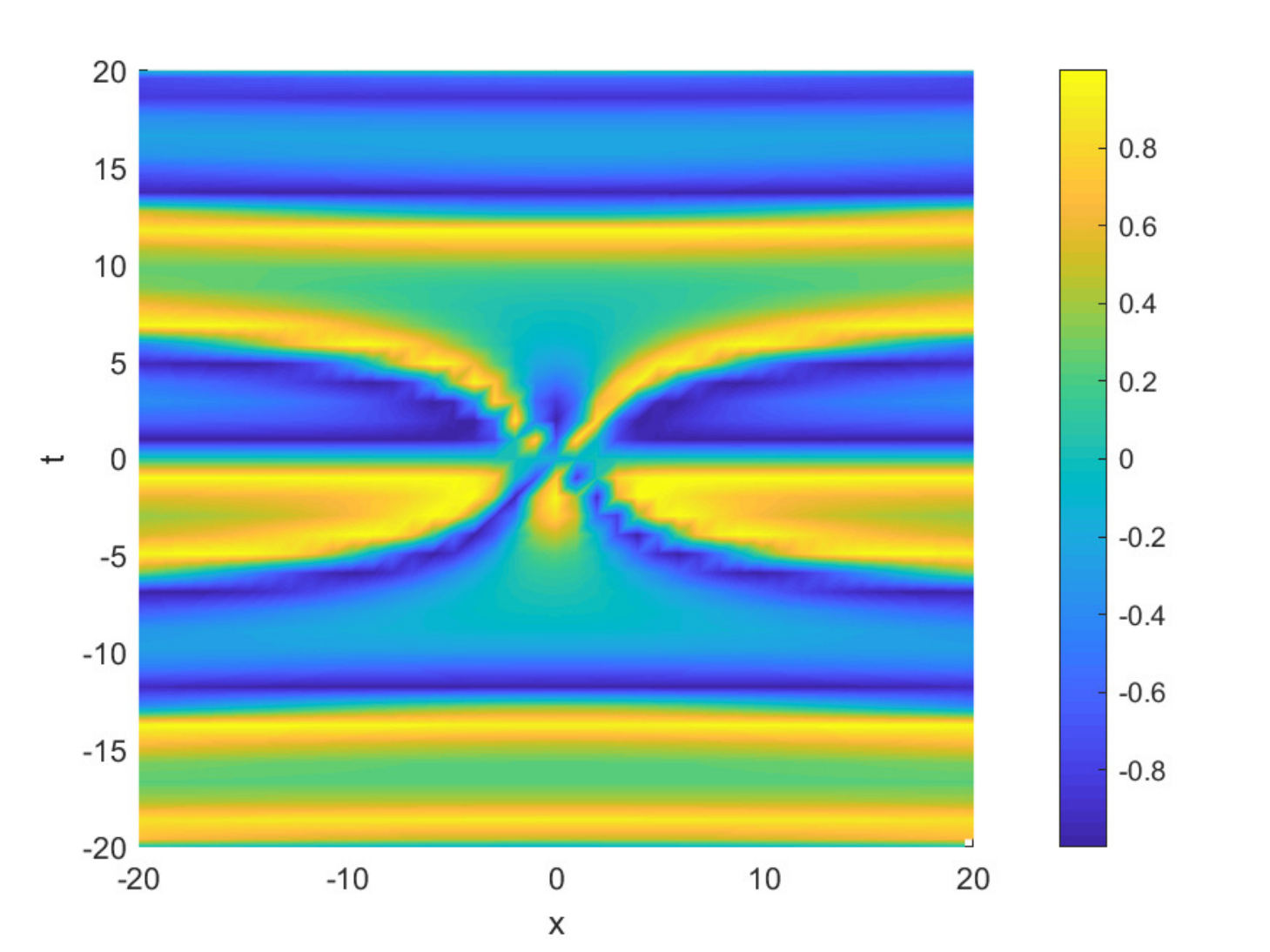}
	\caption{Surface plots of $\sin(u)$ versus $(x,t)$ 
		for rogue waves on the background of librational waves 
		for different values of $k$. 
		Top left: $k=\sin(\frac{\pi}{24})$.
		Top right: $k=\sin(\frac{\pi}{6})$. 
		Bottom left: $k=\sin(\frac{\pi}{3})$.
		%bottom left: $k=\sin(\frac{3\pi}{8})$, 
		Bottom right: $k=\sin(\frac{11\pi}{24})$.}
	% (e) $k=\sin(\frac{7\pi}{24})$, 
	\label{sinmiller}
\end{figure}

Rogue waves on the background of librational waves are displayed on Figure \ref{sinmiller}, where surface plots of $\sin(u)$ are plotted versus $(x,t)$. 
The wave patterns are very similar to the solutions from Appendix D of \cite{milluby}. This confirms that rogue waves on a background of librational waves model defects in the fluxon condensate obtained in \cite{milluby} from the Riemann--Hilbert problem.

\begin{figure}[htb!] 
\includegraphics[width=0.48\textwidth]{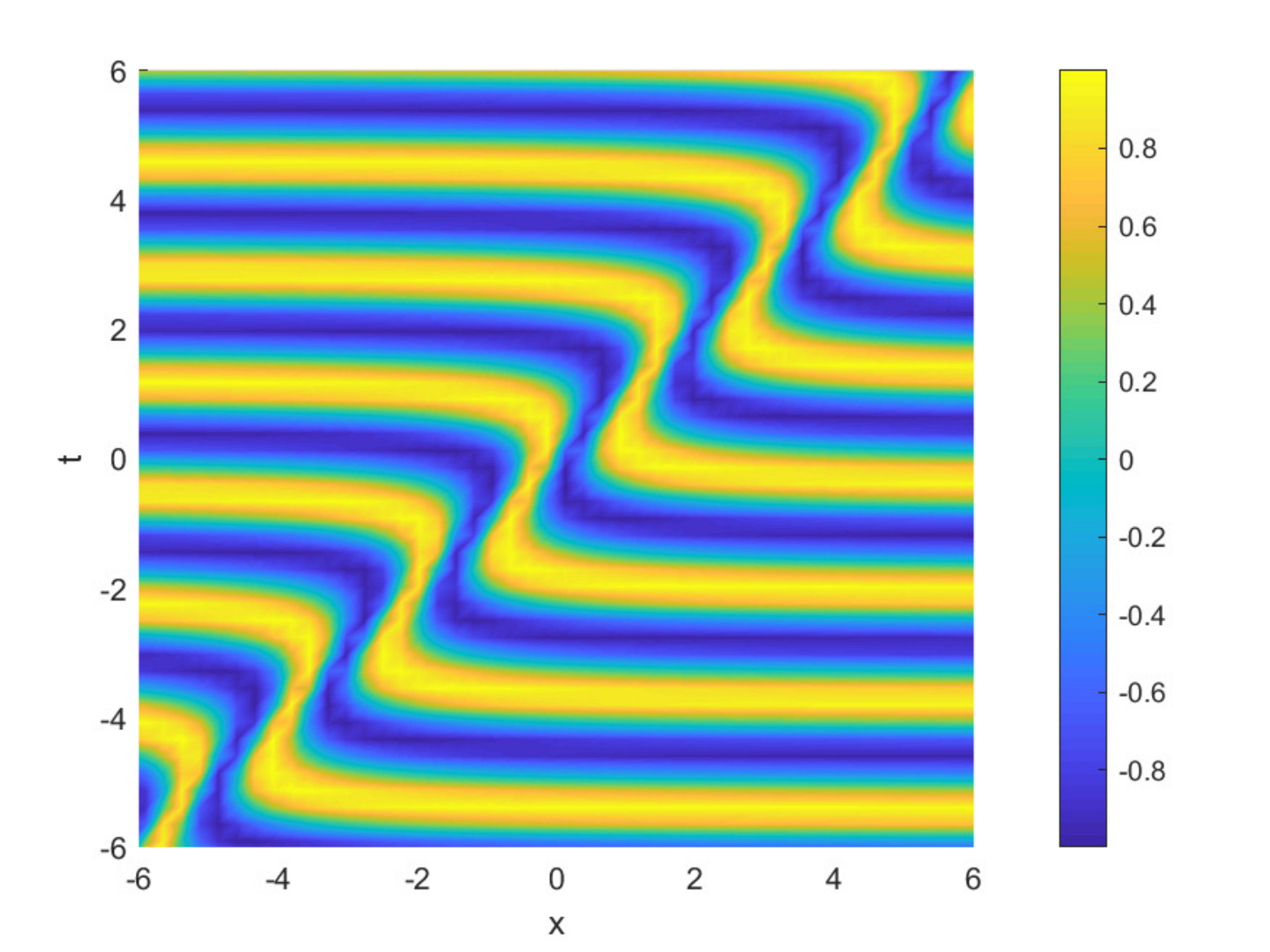}
\includegraphics[width=0.48\textwidth]{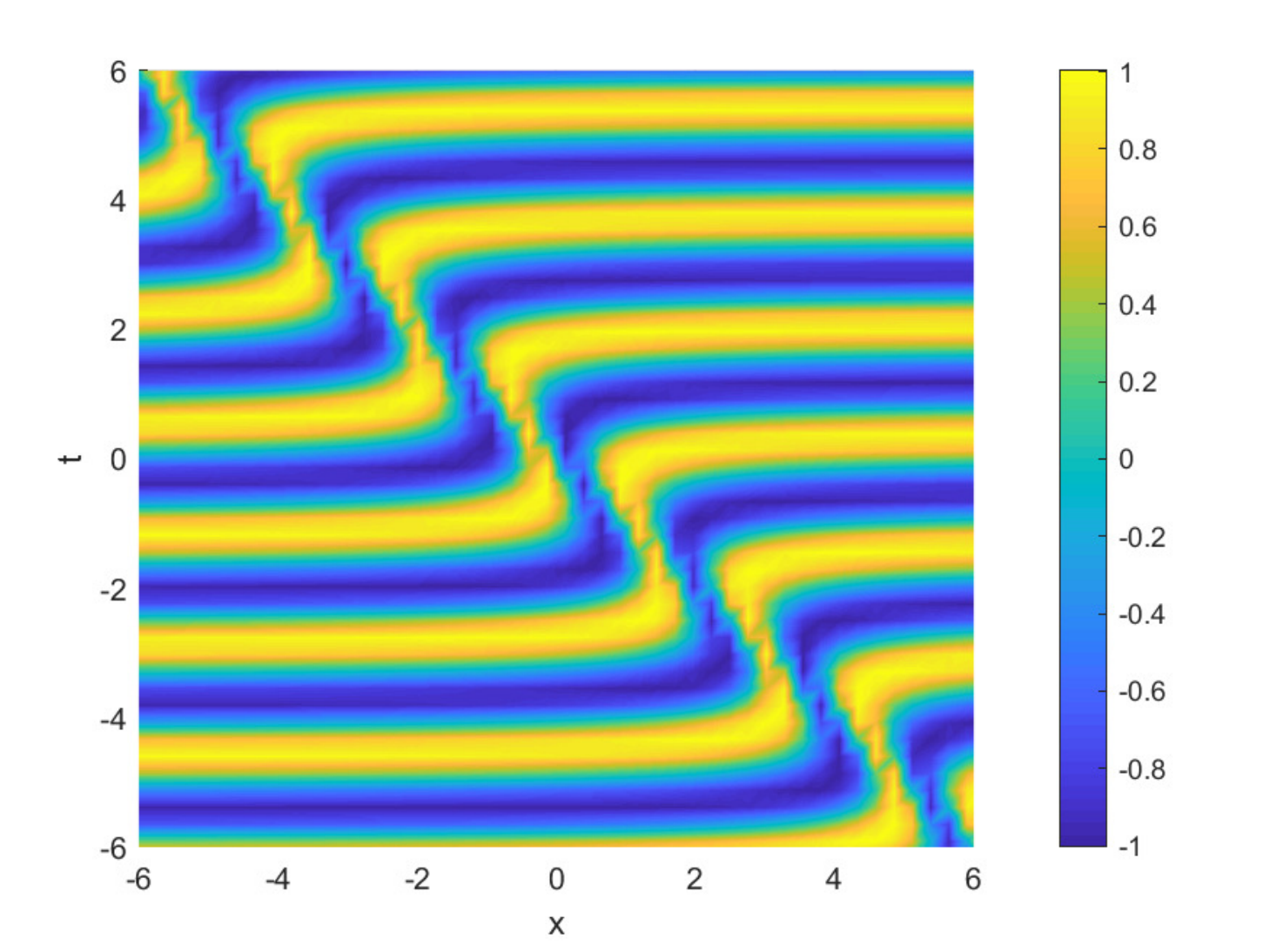} \\
\includegraphics[width=0.48\textwidth]{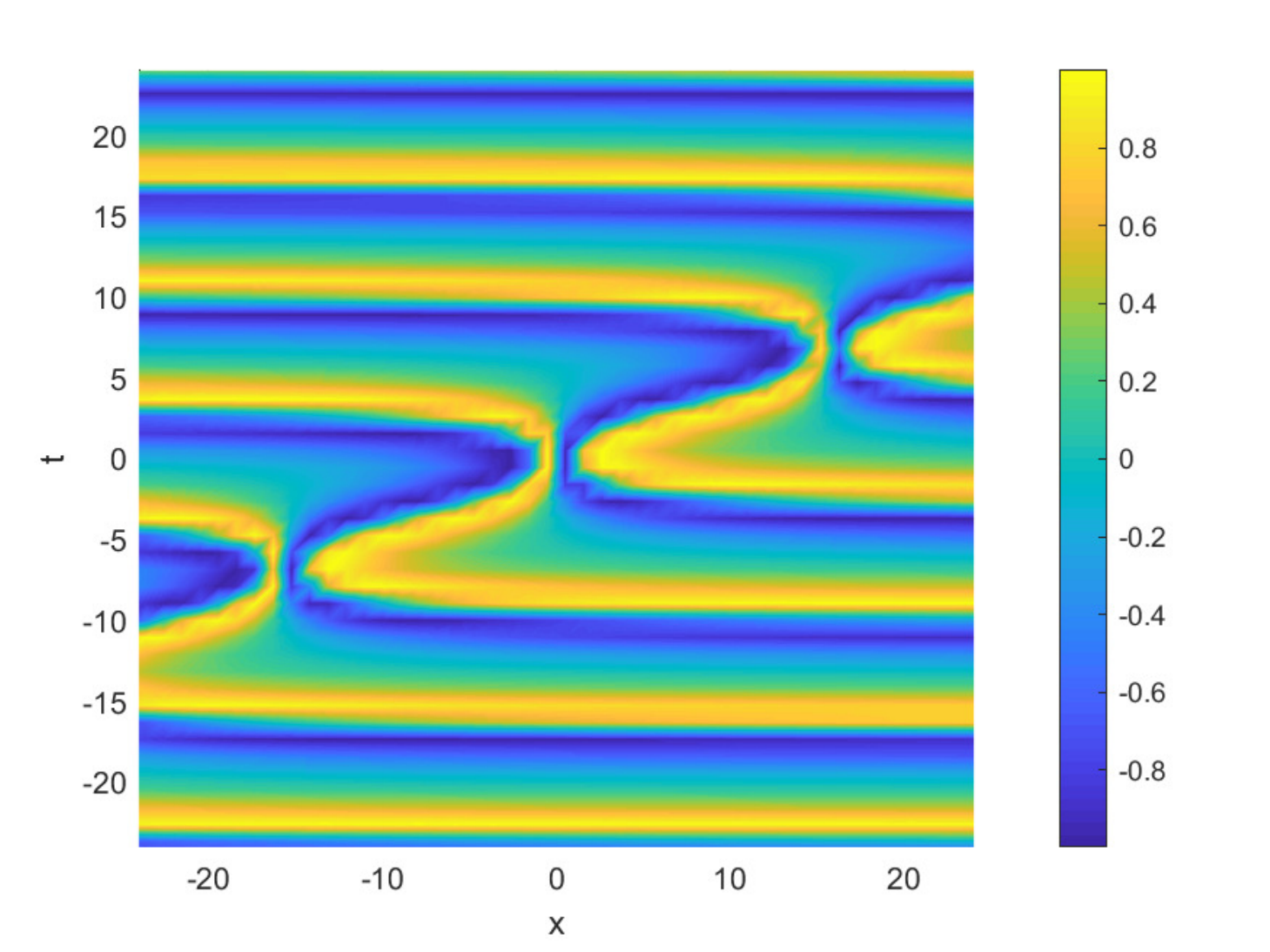}
\includegraphics[width=0.48\textwidth]{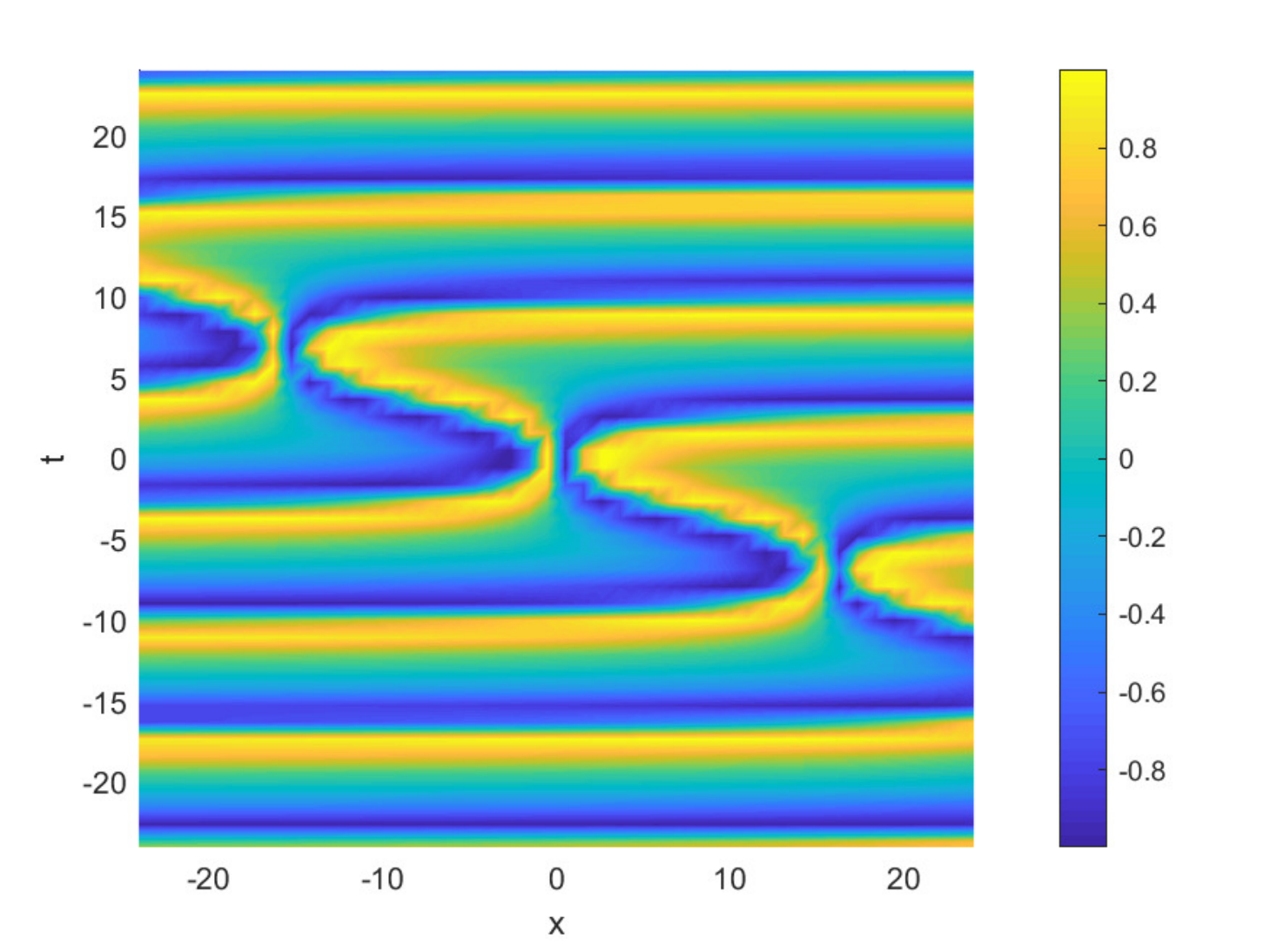}
	\caption{Surface plots of $\sin(u)$ versus $(x,t)$ 
		for kinks (left) and antikinks (right) propagating on the background of rotational waves 	for different values of $k$. 
		Top: $k=\sin(\frac{\pi}{6})$. Bottom: $k=\sin(\frac{11\pi}{24})$.}
	\label{sinRotational}
\end{figure}

Kinks and antikinks propagating on the background of rotational waves are shown on Figure \ref{sinRotational}, where the surface plots of $\sin(u)$ are plotted versus $(x,t)$. The wave patterns appear very similar to the propagation of kinks and antikinks studied for the dynamics of the fluxon condensate 
in the semi-classical limit \cite{Mil1,Mil2}. 

This article is organized as follows. Travelling periodic waves 
of the sine--Gordon equation are expressed by elliptic functions in Section 2. Lax equations are introduced for the sine--Gordon equation in characteristic variables in Section 3. The algebraic method is developed in Section 4, where the bounded periodic eigenfunctions 
in space-time coordinates are explicitly computed for particular 
eigenvalues in the Lax spectrum. 
The new solutions on the background of the rotational (libratitional) waves are constructed in Section 5 (Section 6).  Section 7 concludes the paper 
with the summary.

\section{Travelling periodic waves}

Travelling wave solutions of the sine-Gordon equation (\ref{SG}) are written in the form
$u(x,t)=f(x-ct)$, where $c$ is the wave speed and $f(x) : \mathbb{R} \rightarrow \mathbb{R}$ 
is the wave profile satisfying the following differential equation:
\begin{equation}
\label{travelingwavereduc}
(c^2 -1) f'' + \sin(f) = 0,
\end{equation} 
where the prime corresponds to differentiation in $x$ (after translation to the right by $ct$). 
Superluminal motion corresponds to $c^2 > 1$,  in which case 
the following transformation $f(x) = \hat{f}(\hat{x})$ with $\hat{x} = x/\sqrt{c^2-1}$ results in the dimensionless equation:
\begin{equation}
\label{SGtrav}
\hat{f}'' + \sin(\hat{f}) = 0,
\end{equation}
where the prime now corresponds to differentiation in $\hat{x}$. 
In what follows, we drop hats for simplicity of notations.

The reason why the travelling wave solutions to the sine--Gordon equation (\ref{SG}) 
can be expressed without wave speed $c$ 
is the following Lorentz transformation for $c^2 > 1$ (a similar transformation exists for $c^2 < 1$):
\begin{equation}
\hat{x} = \frac{x-ct}{\sqrt{c^2-1}}, \quad 
\hat{t} = \frac{t-cx}{\sqrt{c^2-1}}, \quad 
\hat{u} = \pi + u, 
\label{lorentz}
\end{equation}  
where $\hat{u}  = \hat{u}(\hat{x},\hat{t})$ satisfies the same sine--Gordon equation (\ref{SG}). The time-independent function 
$\hat{u}(\hat{x},\hat{t}) = \pi + \hat{f}(\hat{x})$ satisfies 
the differential equation (\ref{SGtrav}).

\begin{figure}[htb!]
	\includegraphics[width=0.7\textwidth]{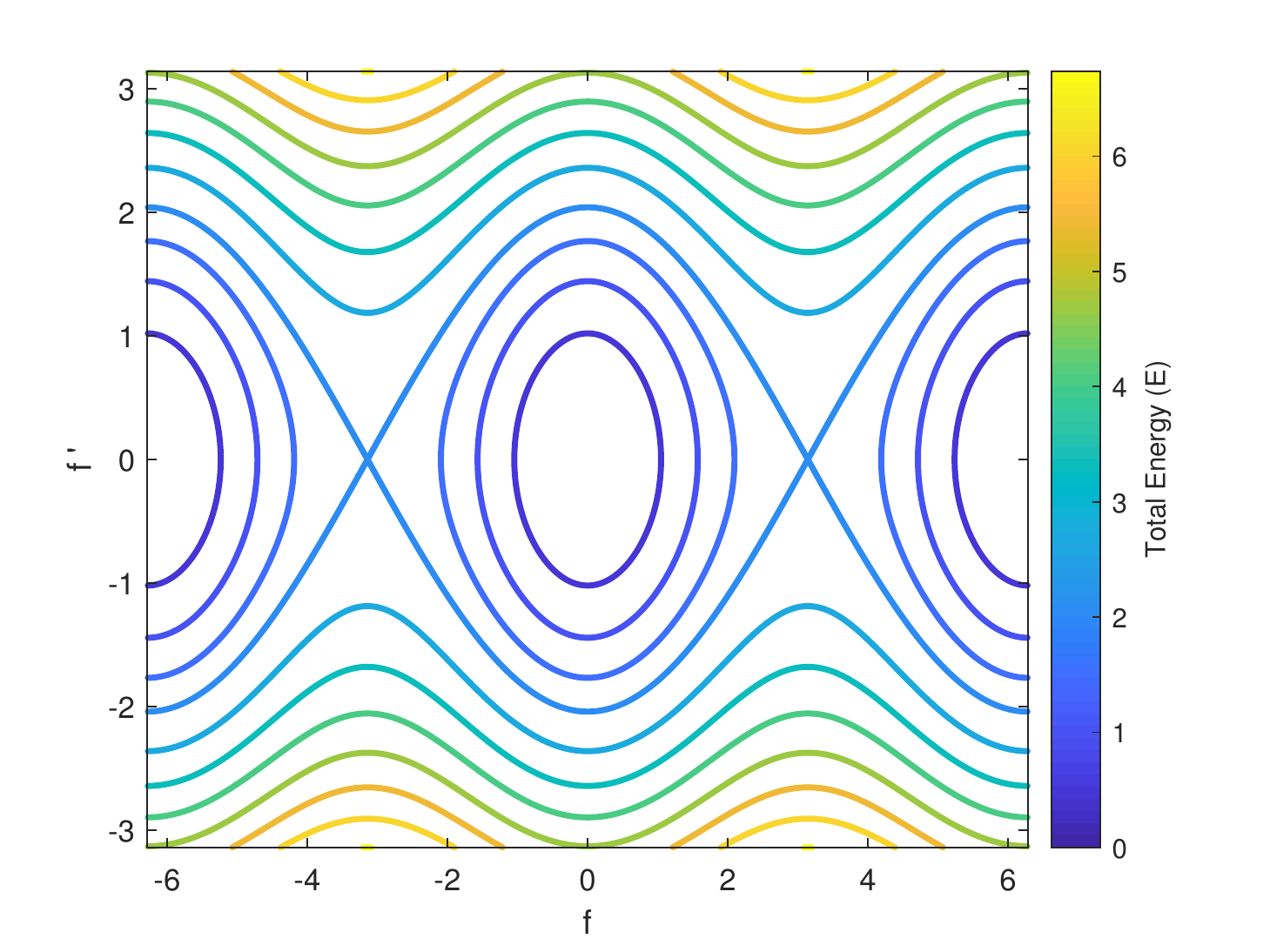}
	\caption{Orbits of the second-order equation (\ref{SGtrav}) on the phase plane $(f,f')$.}
	\label{fig:2}
\end{figure}

The second-order equation (\ref{SGtrav}), where hats are now dropped, is integrable with the first-order invariant:
	\begin{equation} 
	E(f,f') := \frac{1}{2} (f')^2 + 1 - \cos(f).
	\label{energy1} 
	\end{equation} 
It is straightforward to verify that $E(f,f')$ is constant in $x$ along the solutions 
of the second-order equation (\ref{SGtrav}).
The level sets of $E(f,f')$ represent all 
solutions to the differential equation (\ref{SGtrav}) as orbits on the phase plane $(f,f')$. Figure \ref{fig:2} plots the level sets of $E(f,f')$.
There are three different cases for $f\in [-\pi,\pi]$. When $E\in(0,2)$ the level curve is a 
periodic orbit centered around $(0,0)$ which corresponds to librational motion. 
When $E=2$ there are two heteroclinic orbits connecting $(-\pi,0)$ to $(\pi,0)$ which are referred 
to as kinks. Orbits for $E>2$ yields rotational motion. 

Exact analytical solutions for the librational and rotational waves 
are available in terms of Jacobi elliptic functions 
${\rm sn}$, ${\rm cn}$, and ${\rm dn}$. These elliptic functions are derived from 
the inversion of the elliptic integral of the first kind,
\begin{equation*}
z = F(\tau,k) = \int_{0}^{\tau} \frac{dt}{\sqrt{1-k^2\sin^2t}},
\end{equation*}
where $k\in(0,1)$ is the elliptic modulus. The complete elliptic integral 
is defined as $K(k) = F(\frac{\pi}{2},k)$. The first two Jacobi elliptic functions 
are defined by $\textrm{sn}(z,k) = \sin \tau$ and $\textrm{cn}(z,k) = \cos \tau$ such that  
\begin{equation}
\textrm{sn}^2(z,k) + \textrm{cn}^2(z,k)  = 1. 
\label{Jid1}
\end{equation}  
These functions  are smooth, sign-indefinite, and periodic with the period $4K(k)$. The third Jacobi elliptic function is defined from the quadratic formula  
\begin{equation}
\textrm{dn}^2(z,k) + k^2\textrm{sn}^2(z,k)  = 1. 
\label{Jid2}
\end{equation} 
The function ${\rm dn}(z,k)$ is given by the positive square root of (\ref{Jid2}), so that it is smooth, positive, and periodic with the period $2 K(k)$. The Jacobi elliptic functions are related by the derivatives:
\begin{equation} 
\label{der-Jac}
\left\{ \begin{array}{l}
\frac{d}{dz} \ \textrm{sn}(z,k) = \textrm{cn}(z,k) \ \textrm{dn}(z,k),\\
\frac{d}{dz} \ \textrm{cn}(z,k) = -\textrm{sn}(z,k) \ \textrm{dn}(z,k), \\ 
\frac{d}{dz} \ \textrm{dn}(z,k) = -k^2 \textrm{sn}(z,k) \ \textrm{cn}(z,k).
\end{array} \right.
\end{equation} 

For $E \in (0,2)$, the librational waves of the first-order invariant (\ref{energy1}) are given up to an arbitrary translation in $x$ by 
\begin{equation}
\label{libr-wave}
\left\{ \begin{array}{cl}
	\cos(f) & = 1 - 2 k^2 {\rm sn}^2(x,k), \\
	\sin(f) & = 2 k {\rm sn}(x,k) {\rm dn}(x,k) \\
	f' & = 2 k {\rm cn}(x,k),
	\end{array} \right.
\end{equation}
where $E = 2 k^2 \in (0,2)$. In order to verify the validity of (\ref{libr-wave}), we note that the first-order invariant (\ref{energy1}) is satisfied due to (\ref{Jid1}), the trigonometric identity is satisfied due to (\ref{Jid2}), 
and the derivative of $\cos(f)$ and $\sin(f)$ are consistent due to (\ref{der-Jac}). 
The period of the librational waves (\ref{libr-wave}) is $L = 4 K(k)$.

For $E \in (2,\infty)$, the rotational waves of the first-order invariant (\ref{energy1}) are given up to an arbitrary translation in $x$ by 
\begin{equation}
\label{rot-wave}
\left\{ \begin{array}{cl}
\cos(f) & = 1 - 2 {\rm sn}^2(k^{-1} x,k), \\
\sin(f) & = \pm 2 {\rm sn}(k^{-1}x,k) {\rm cn}(k^{-1}x,k) \\
f' & = \pm 2 k^{-1} {\rm dn}(k^{-1}x,k),
\end{array} \right.
\end{equation}
where $E = 2 k^{-2} \in (2,\infty)$ and the upper/lower sign corresponds to the orbit 
in the upper/lower half plane on Fig. \ref{fig:2}. Again, 
the first-order invariant (\ref{energy1}) is satisfied due to (\ref{Jid2}), the trigonometric identity is satisfied due to (\ref{Jid1}), 
and the derivative of $\cos(f)$ and $\sin(f)$ are consistent due to (\ref{der-Jac}).
The period of the rotational waves (\ref{rot-wave}) is $L = 2 k K(k)$.

\section{Lax equations in characteristic coordinates}

Lax equations for the sine--Gordon equation (\ref{SG}) are rather combursome \cite{Mil2,DMS}. Therefore, we adopt the following characteristic coordinates:
\begin{align}
\label{changeofvar}
\xi = \frac{1}{2}(x+t), \qquad
\eta = \frac{1}{2}(x-t).
\end{align}  
The sine--Gordon equation (\ref{SG}) can be written in a simpler form:
\begin{equation}
u_{\xi\eta} = \sin(u),
\label{sine-gordon-en} 
\end{equation}
where $u = u(\xi,\eta)$. 
The travelling periodic wave is now given by 
$u(\xi,\eta) = \hat{f}(\xi-\eta)$, where 
$\hat{f}(\xi) : \mathbb{R} \rightarrow \mathbb{R}$ satisfies 
the second-order equation (\ref{SGtrav}), where 
the prime represents the derivative with respect to $\hat{x} = \xi - \eta = t$. 
Note that $t$ and $\hat{x}$ are equivalent due to the Lorenz 
transformation (\ref{lorentz}).

Lax equations for the sine-Gordon equation 
in characteristic coordinates (\ref{sine-gordon-en}) 
are given by the following system:
\begin{equation}
\frac{\partial}{\partial \xi} 
\begin{bmatrix} p\\q \end{bmatrix} 
= \frac{1}{2} 
\begin{bmatrix} \lambda & -u_{\xi}\\u_{\xi} &-\lambda \end{bmatrix}
\begin{bmatrix} p\\q \end{bmatrix} \label{linearlaxnonlab}
\end{equation} 
and
\begin{equation}
\frac{\partial}{\partial \eta} 
\begin{bmatrix} p\\q \end{bmatrix} 
= \frac{1}{2\lambda} 
\begin{bmatrix} \cos(u) & \sin(u)\\ \sin(u) &-\cos(u) \end{bmatrix}
\begin{bmatrix} p\\q \end{bmatrix}, \label{linearlaxnonlab2}
\end{equation}
where $\lambda\in \mathbb{C}$ is the spectral parameter and 
$\chi := (p,q)^T$ is an eigenfunction written in variables
$(\xi,\eta)$. Validity of the sine--Gordon equation 
(\ref{sine-gordon-en}) as the compatibility condition 
$\chi_{\xi \eta} = \chi_{\eta \xi}$ can be checked by 
direct differentiation \cite{AKNS}. 
The first equation (\ref{linearlaxnonlab}) 
is referred to as the AKNS spectral problem with the potential $w := - u_{\xi}$. 

When $w = -\hat{f}'(\hat{x})$ is a travelling periodic wave 
with the fundamental period $L$, 
the AKNS spectral problem determines the Lax spectrum 
in $L^2(\mathbb{R})$ as the set of all admissible values 
of $\lambda$ for which $\chi \in L^{\infty}(\mathbb{R})$.  
By Floquet theorem, bounded solutions of the linear equation 
(\ref{linearlaxnonlab}) can be represented in the form  
\begin{equation}\label{floq}
\chi(\xi,\eta) = \phi(\xi-\eta) e^{i\mu (\xi - \eta) + \Omega \eta},
\end{equation}
where $\phi$ is $L$-periodic, $\mu$ is defined in the fundamental 
region $[-\frac{\pi}{L},\frac{\pi}{L}]$, and $\Omega$ is a new spectral 
parameter arising in the separation of variables in the second 
Lax equation (\ref{linearlaxnonlab2}) \cite{DMS}. The admissible values of $\lambda$ in $\mathbb{C}$ are defined by periodic solutions of the 
following eigenvalue problem:
\begin{equation}\label{floq2}
\begin{bmatrix}
2\frac{d}{d\hat{x}} + 2i\mu & \hat{f}'(\hat{x})\\
\hat{f}'(\hat{x}) & -2\frac{d}{d\hat{x}} - 2i\mu
\end{bmatrix} 
\phi =  \lambda \phi,
\end{equation}
where $\hat{x}:= \xi - \eta$ and $\mu \in [-\frac{\pi}{L},\frac{\pi}{L}]$. 
The spectral parameter $\Omega$ determines an eigenvalue of the spectral stability problem for the travelling periodic wave evolving with respect to the coordintae $\eta$ (see Theorem 5.1 in \cite{DMS} for spectral stability of the travelling periodic wave evolving with respect to the time variable $t$). 
Compared to \cite{DMS}, we will not explore the spectral stability 
of travelling periodic waves but will construct solutions to the Lax equations (\ref{linearlaxnonlab})  and (\ref{linearlaxnonlab2}) which correspond to $\Omega = 0$. Such eigenfunctions $\chi$ are bounded in both $\xi$ and $\eta$, 
hence in the space-time coordinates $(x,t)$.

\section{Algebraic method}

The purpose of the algebraic method is to relate solutions of 
the nonlinear integrable equation and solutions of the associated linear 
Lax equations in order to obtain an explicit expression for the particular eigenvalues of the Lax spectrum. These eigenvalues correspond to bounded eigenfunctions in the space-time coordinates. 
Our presentation of the algebraic method follows closely to \cite{CPkdv,CPgardner} devoted to the mKdV equation because 
the AKNS spectral problem (\ref{linearlaxnonlab}) is identical 
with the potential $w(\hat{x}) := -\hat{f}'(\hat{x})$, 
where $\hat{x} = \xi-\eta = t$. As previously mentioned, we will drop hats for simplicity of notations.

Assume that $(p_1,q_1)$ is a solution to the AKNS spectral problem (\ref{linearlaxnonlab}) for a 
fixed value of $\lambda = \lambda_1$. Assume that 
the solution $u = u(\xi,\eta)$ to the sine--Gordon equation (\ref{sine-gordon-en}) is related 
to the squared eigenfunctions by 
\begin{equation}\label{squareeigen}
-u_{\xi} = p_1^2 + q_1^2 .
\end{equation} 
The linear equation (\ref{linearlaxnonlab}) with the constraint (\ref{squareeigen}) becomes a nonlinear Hamiltonian system with Hamiltonian 
\begin{equation}\label{hammmy}
H(p_1,q_1) = \lambda_1 p_1 q_1 + \frac{1}{4}(p_1^2+q_1^2)^2,
\end{equation}
so that 
\begin{equation}
\frac{\partial}{\partial \xi} 
\begin{bmatrix} p_1 \\ q_1 \end{bmatrix} 
= \frac{1}{2} 
\begin{bmatrix} 0 & 1 \\ -1 & 0 \end{bmatrix}
\begin{bmatrix} \frac{\partial H}{\partial p_1} \\ \frac{\partial H}{\partial q_1} \end{bmatrix}. \label{nonlinearlax}
\end{equation}
Let us denote the constant value of $H(p_1,q_1)$ at the solutions 
of (\ref{nonlinearlax}) by $H_0$ so that 
\begin{equation}
\label{hammmy-F}
\lambda_1 p_1 q_1 = H_0 - \frac{1}{4} (f')^2.
\end{equation}
Recall that $u(\xi,\eta) = f(x)$ solves the second-order equation 
\begin{equation}\label{sinft}
f'' + \sin(f) = 0,
\end{equation}
derivative of which yields 	
	\begin{equation}\label{cosft}
	f''' + \cos(f)f' = 0.
	\end{equation}
	Comparing (\ref{cosft}) with (\ref{energy1}) and eliminating  
	$\cos(f)$ produces the third-order equation 
	\begin{equation}\label{algebraic2}
	f''' = f' (E-1) - \frac{1}{2}(f')^3,
	\end{equation}
	where $E$ is constant. 
	
Differentiating the constraint (\ref{squareeigen}) twice and using 
(\ref{nonlinearlax}) gives 
	\begin{equation}
	\label{useful123} 
-f'' = \lambda_1 (p_1^2-q_1^2)
\end{equation}
and
\begin{equation}
\label{useful123-add} 
f''' = \lambda_1^2 f' + 2\lambda_1 f' p_1q_1.
\end{equation}
Substituting (\ref{hammmy-F}) for $\lambda_1 p_1 q_1$ into (\ref{useful123-add}) yields 	
	\begin{equation}\label{algebraic1}
	f''' = \lambda_1^2 f' +2 H_0 f' - \frac{1}{2} (f')^3.
	\end{equation} 
		Comparing (\ref{algebraic1}) with (\ref{algebraic2}) gives 
		the following relation:
	\begin{equation}\label{FdaL1}
E = \lambda_1^2  + 2 H_0 + 1.
\end{equation} 

In order to determine the explicit formula for $\lambda_1$ in terms of $E$, 
we shall integrate the nonlinear system (\ref{nonlinearlax}) by using the Lax equation:
\begin{equation}\label{laxx2} 
2\frac{\partial}{\partial \xi} W(\lambda) = Q(\lambda)W(\lambda) - W(\lambda)Q(\lambda),
\end{equation} 
where
\begin{equation}
Q(\lambda) = 
\begin{bmatrix}
\lambda & p_1^2 + q_1^2\\-p_1^2-q_1^2 &-\lambda
\end{bmatrix}, \quad 
W(\lambda) = 
\begin{bmatrix}
W_{11}(\lambda) & W_{12}(\lambda)\\ W_{12}(-\lambda) &-W_{11}(-\lambda)
\end{bmatrix}
\end{equation} 
with
\begin{equation}
\label{Wpq} 
W_{11}(\lambda) = 1 - \frac{p_1q_1}{\lambda-\lambda_1} + \frac{p_1q_1}{\lambda 
	+ 	\lambda_1}, \quad 
W_{12}(\lambda) = \frac{p_1^2}{\lambda-\lambda_1} + \frac{q_1^2}{\lambda + 
	\lambda_1}.
\end{equation}
Substituting (\ref{squareeigen}), (\ref{hammmy-F}), and (\ref{useful123}) into (\ref{Wpq}) yields the following expressions:
\begin{equation}
\label{Wu}
W_{11}(\lambda) = 1 - \frac{4 H_0-(f')^2}{2(\lambda^2 -\lambda_{1}^2)}, \quad 
W_{12}(\lambda) = \frac{-\lambda f' - f''}{\lambda^2 - \lambda_{1}^2}.
\end{equation}
The determinant of $W(\lambda)$ is computed from (\ref{Wpq}) as 
	\begin{align*}
	\det[W(\lambda)] &= -[W_{11}(\lambda)]^2 - W_{12}(\lambda)W_{12}(-\lambda) \\ 
	&= -1 + \frac{4\lambda_1 p_1q_1 + (p_1^2+q_1^2)^2}{\lambda^2 - 
		\lambda_1^2}\\
	&= -1 + \frac{4H_0}{\lambda^2 - \lambda_1^2},
	\end{align*}
	where we have used (\ref{hammmy-F}). Hence, $\det[W(\lambda)]$ only admits simple poles at $\lambda = \pm \lambda_1$. On the other hand, the determinant of $W(\lambda)$ is computed from (\ref{Wu}) as
	\begin{align*}
	det[W(\lambda)] &= -1 + \frac{4 H_0-(f')^2}{\lambda^2 -\lambda_{1}^2} 
	+ \frac{(\lambda^2 + 2 H_0)(f')^2 - (f'')^2 - 4 H_0^2 - \frac{1}{4} (f')^4}{(\lambda^2 -\lambda_{1}^2)^2} \\
	&= -1 + \frac{4 H_0}{\lambda^2 -\lambda_{1}^2} 
	+ \frac{4 (\lambda_1^2 + 2 H_0)(f')^2 - 4 (f'')^2 - 16 H_0^2 - (f')^4}{4 (\lambda^2 -\lambda_{1}^2)^2}.
	\end{align*}
Comparison of these two equivalent expressions yields the constraint:	
	\begin{equation}\label{alg4}
(\lambda_1^2 + 2 H_0)(f')^2 - (f'')^2 - 4 H_0^2 - \frac{1}{4} (f')^4 = 0.
	\end{equation}
By the fundamental trigonometric identity, we obtain from (\ref{energy1}) and
(\ref{sinft}):
	\begin{equation}\label{alg5}
	1 = \sin^2(f) + \cos^2(f) = (f'')^2 + 
	\frac{1}{4} (f')^2 + (1 - E) (f')^2 + (1-E)^2.
	\end{equation}
Comparing (\ref{alg4}) and (\ref{alg5}) yields the relation 
	\begin{equation}\label{FdLa2}
4 H_0^2 = E(E-2), 
\end{equation}
in addition to (\ref{FdaL1}). Expressing $H_0$ from (\ref{FdLa2}) and substituting into (\ref{FdaL1}) yield admissible values of $\lambda_1$ by 
	\begin{align} \label{redots}
	\lambda_1^2 = E - 1 \mp \sqrt{E(E-2)}, 
	\end{align}
where the plus and minus sign correspond to the two roots in 
	\begin{align}\label{rdots2} 
	2 H_0 = \pm \sqrt{E(E-2)}.  
	\end{align}
	
For the rotational waves (\ref{rot-wave}), we have $E = 2/k^2$, so that one can extract the square root 
from (\ref{redots}) and obtain two real pairs of admissible values $\pm \lambda_1$ with 
\begin{align}
\lambda_{1} = \frac{1 \mp \sqrt{1-k^2}}{k}, 
\label{Rdots}
\end{align}
where the plus and minus signs correspond to the signs in
\begin{align}\label{Rdots-F} 
H_0 = \pm \frac{\sqrt{1-k^2}}{k^2}. 
\end{align}

For the libratitional waves (\ref{libr-wave}), we have $E = 2 k^2$ so that 
one can again extract the square root 
from (\ref{redots}) and obtain a complex quadruplet of admissible values 
$\{\pm \lambda_1, \pm \bar{\lambda}_1 \}$ with
\begin{align}
\lambda_{1} = k + i\sqrt{1-k^2},
\label{Ldots}
\end{align}
where the unique $\lambda_1$ is located in the first quadrant of the complex plane. This eigenvalue corresponds to the choice in 
\begin{equation}
\label{Ldots-F}
H_0 = - ik \sqrt{1-k^2}.
\end{equation}
 
\begin{figure}[htb!]
	\includegraphics[width=0.45\textwidth]{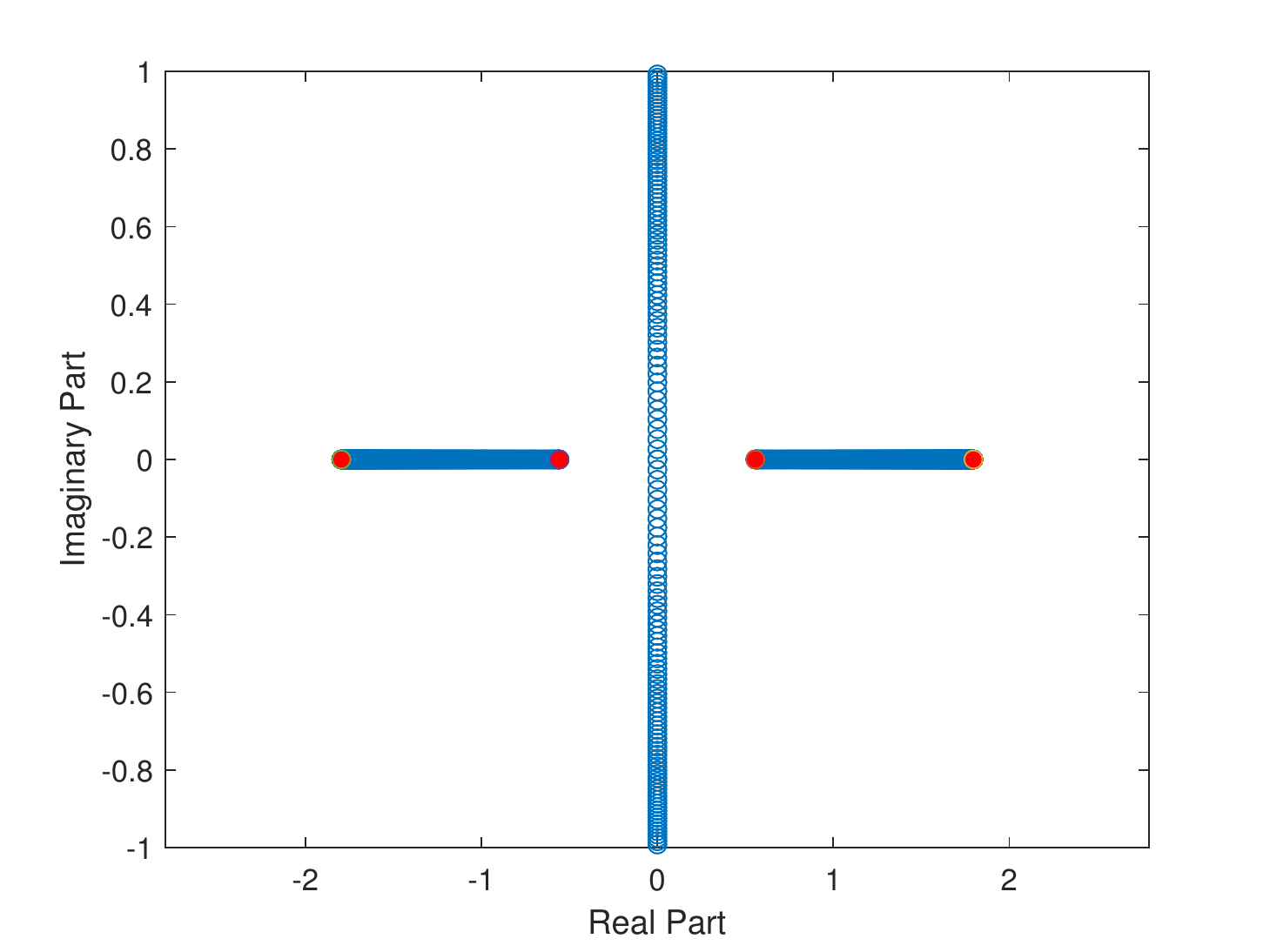}
	\includegraphics[width=0.45\textwidth]{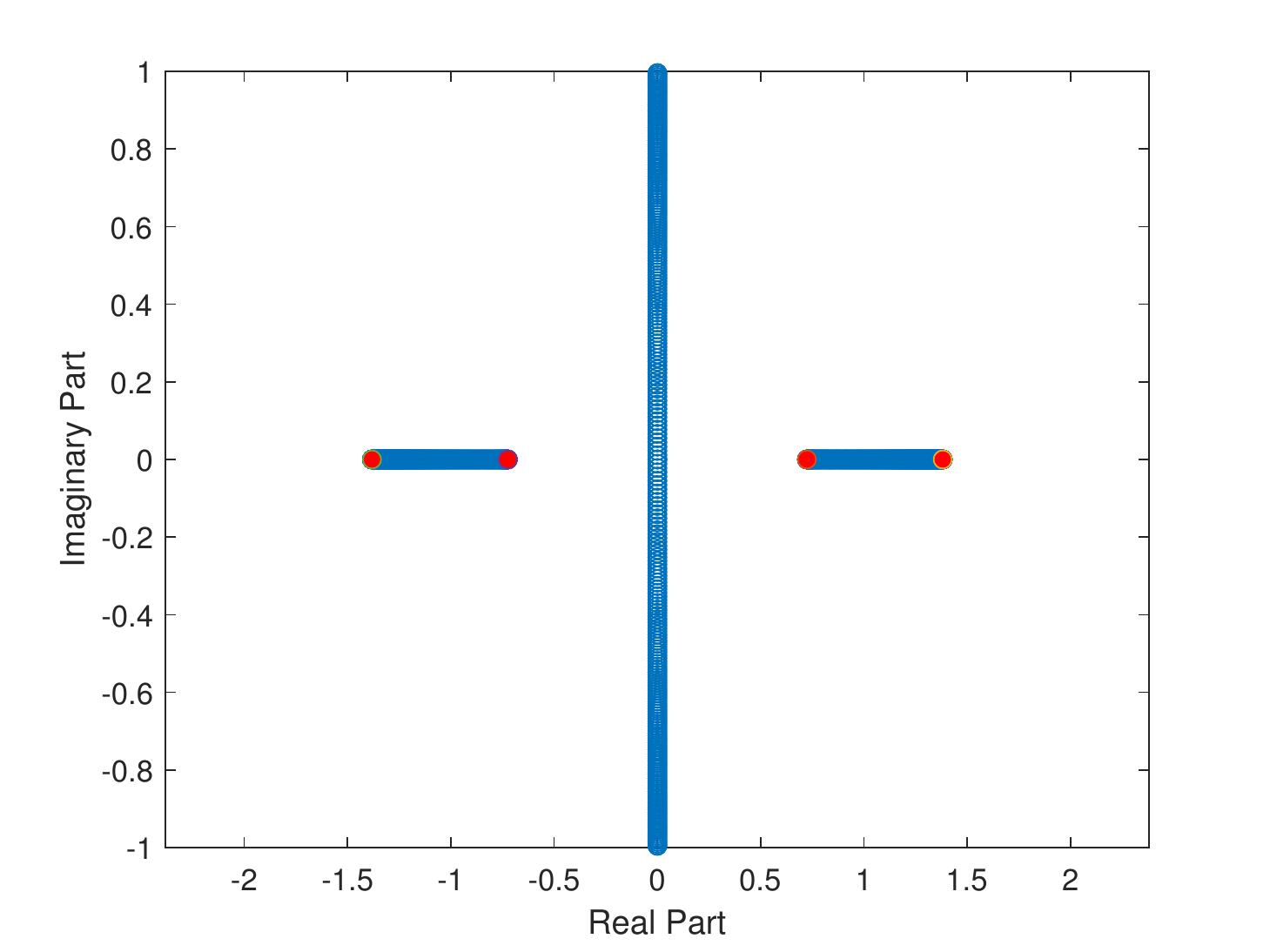}\\
	\includegraphics[width=0.45\textwidth]{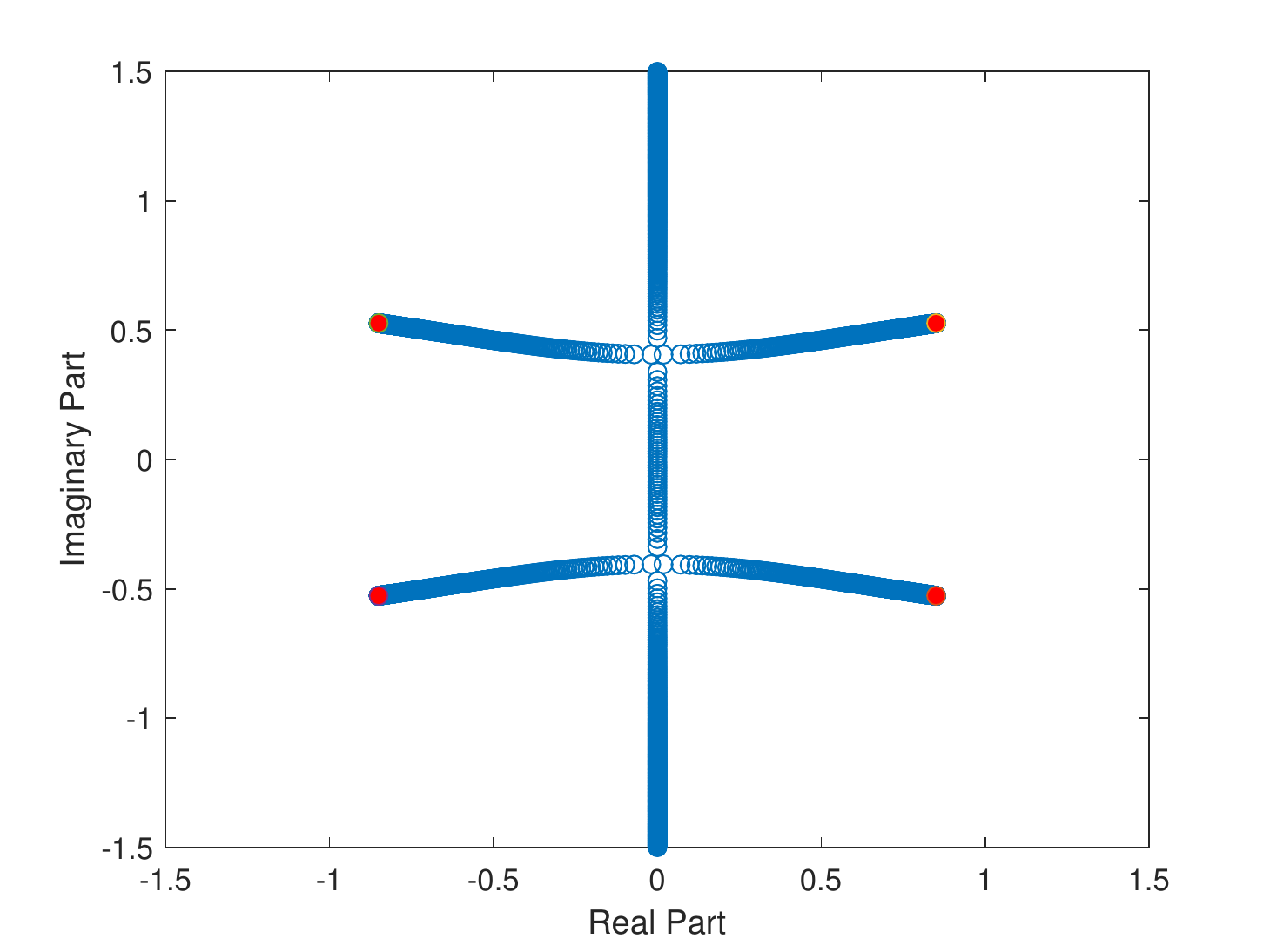}
	\includegraphics[width=0.45\textwidth]{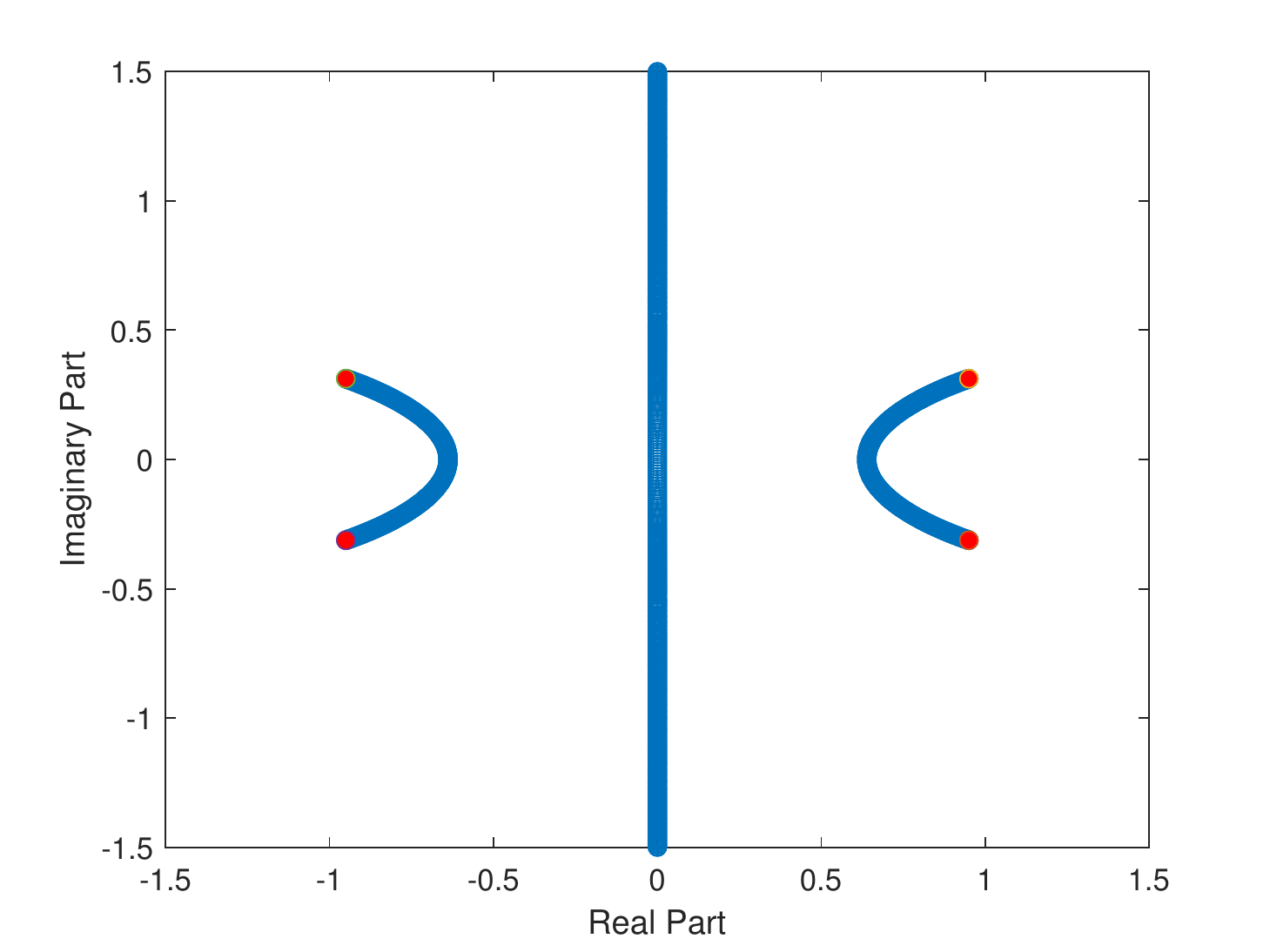}
	\caption{The Lax spectrum of (\ref{linearlaxnonlab}) associated with the 
		rotational (top) and librational (bottom) waves for $k=0.85$ 
		(left) and $k=0.95$ (right). Red dots represent eigenvalues 
		(\ref{Rdots}) and (\ref{Ldots}).}
	\label{laxpsec1}
\end{figure}

We approximate numerically the Lax spectrum of the AKNS spectral problem 
(\ref{linearlaxnonlab}) by using the Floquet theorem and converting the spectral problem to the form (\ref{floq2}). By using discretization of the spatial domain 
$[0,L]$ and the range of the $\mu$ values in $[-\frac{\pi}{L},\frac{\pi}{L}]$ we reduce (\ref{floq2}) to the matrix eigenvalue problem for each $\mu$, 
this problem is handled using Matlab's eig() function. The derivative operator 
$\frac{d}{d \hat{x}}$ is replaced with the $12^{th}$ order finite difference matrix to ensure high accuracy of computations. The union of each set of 
eigenvalues associated for each $\mu$ defines the Lax spectrum.

Figure \ref{laxpsec1} shows the numerically constructed Lax 
spectra for the rotational and librational waves using 
certain values of $k$. The end points of the spectral bands outside 
$i \mathbb{R}$ correspond to the eigenvalues (\ref{Rdots}) and (\ref{Ldots}). 

Lax spectra on Figure \ref{laxpsec1} correspond to the AKNS spectral problem (\ref{linearlaxnonlab}) for the sine--Gordon equation in characteristic variables $(\xi,\eta)$. The location of the Lax spectrum in space-time coordinates $(x,t)$ is different because the bounded eigenfunctions 
in $\xi$ are located at different values of $\lambda \in \mathbb{R}$ compared to bounded functions in $x = \xi + \eta$. Nevertheless, the eigenvalues 
(\ref{Rdots}) and (\ref{Ldots}) belong to the Lax spectrum in $(x,t)$ 
because the corresponding eigenfunctions are bounded and periodic both in $x$ and $t$. The same eigenvalues are shown by crosses on Fig. 7 in \cite{DMS}, from which it is clear that the eigenvalues (\ref{Rdots}) and (\ref{Ldots}) do not appear as the end points of the Lax spectrum in the space-time coordinates $(x,t)$.

\section{New solutions on the background of rotational waves}

Let $(p,q)$ be a solution to the linear equations (\ref{linearlaxnonlab}) and (\ref{linearlaxnonlab2}) for a fixed value of $\lambda$ and for the solution $u = u(\xi,\eta)$ of the sine--Gordon equation (\ref{sine-gordon-en}). As is shown in \cite{CPkdv}, the new solution $\hat{u}= \hat{u}(\xi,\eta)$ 
to the sine--Gordon equation is given by the 
one-fold Darboux transformation: 
\begin{equation}
\label{ODT}
\hat{w} = w + \frac{4\lambda p q}{p^2 + q^2},
\end{equation}
where $w := -u_{\xi}$ and $\hat{w} := - \hat{u}_{\xi}$. 
If $u = f(\xi-\eta)$ is the rotational wave given by (\ref{rot-wave}) 
with $x = \xi - \eta$ 
and $\lambda = \lambda_1$ is given by the algebraic method 
with the eigenfunction $(p,q) = (p_1,q_1)$ satisfying 
(\ref{squareeigen}) and (\ref{hammmy-F}), 
the one-fold Darboux transformation (\ref{ODT}) yields
\begin{equation}
\hat{w} = w + \frac{4\lambda_1 p_1 q_1}{p_1^2 + q_1^2} = 
w + \frac{4 H_0 - w^2}{w} = \frac{4 H_0}{w}. \label{simpleODT}
\end{equation}
Since $w = -f'$ is given by (\ref{rot-wave}) and 
$H_0$ is given by (\ref{Rdots-F}), we obtain up to the sign 
changes:
\begin{equation}
\hat{w} = \pm \frac{2 k^{-1} \sqrt{1-k^2}}{{\rm dn}(k^{-1} x;k)} 
= \pm 2 k^{-1} {\rm dn}(k^{-1} x + K(k);k).
\label{rottrans}
\end{equation}
The new solution (\ref{rottrans}) is just a half-period translated and reflected version of the rotational wave, which is periodic with the period $L = 2 k K(k)$.

In order to construct a new solution to the sine--Gordon equation 
on the background of the rotational wave (\ref{rot-wave}), we are looking for the second, linear independent solution to 
the linear equations (\ref{linearlaxnonlab}) and (\ref{linearlaxnonlab2}) 
with the same value of $\lambda = \lambda_1$. We will define the 
second solution in the same form as is used in \cite{CPgardner}:
\begin{equation}\label{hatform}
\hat{p}_1 = p_1\phi_R - \frac{q_1}{p_1^{2}+q_1^{2}}, \quad
\hat{q}_1 = q_1\phi_R + \frac{p_1}{p_1^{2}+q_1^{2}}, 
\end{equation}
where the function $\phi_R = \phi_R(\xi,\eta)$ satisfies the system of scalar equations:
\begin{equation}\label{phE-prelim}
\frac{\partial \phi_R}{\partial \xi} = -\frac{2 \lambda_1 p_1 q_1}{(p_1^2 + q_1^2)^2}, \quad 
\lambda_1 \frac{\partial \phi_R}{\partial \eta} = \frac{(p_1^2 - q_1^2)\sin(f)  - 2 p_1 q_1 \cos(f)}{(p_1^2 + q_1^2)^2}.
\end{equation}
The representation (\ref{hatform}) is non-singular for the rotational waves 
because $w = p_1^2 + q_1^2 = -f'$ has the sign-definite $f'$ in (\ref{rot-wave}). 
%The representation (\ref{phE}) was derived in \cite{Robert-thesis}. 
As we prove below, the exact expression for $\phi_R$ is given by 
\begin{equation}
\phi_R(\xi,\eta) = C + \frac{1}{2} (\xi + \eta) - 2 H_0 \int_{0}^{\xi - 
	\eta} \frac{dx}{(f')^2},
\label{growthrotexpl}
\end{equation}
where $C$ is an arbitrary constant of integration. Indeed, 
by using (\ref{squareeigen}), (\ref{hammmy-F}), and (\ref{useful123}) 
we rewrite (\ref{phE-prelim}) in the form:
\begin{equation}\label{phE}
\frac{\partial \phi_R}{\partial \xi} = \frac{1}{2} - \frac{2H_0}{(f')^2}, \quad 
\lambda_1^2 \frac{\partial \phi_R}{\partial \eta} = -\frac{2 f'' \sin(f)  +  (4 H_0 - (f')^2) \cos(f)}{2 (f')^2}.
\end{equation}
By using  (\ref{energy1}), (\ref{sinft}), (\ref{FdaL1}), 
and (\ref{alg4}), the second equation of system (\ref{phE}) 
is simplified to
\begin{align}
\frac{\partial \phi_R}{\partial \eta} = \frac{1}{2} + 
\frac{2H_0}{(f')^2},
\end{align} 
which implies (\ref{growthrotexpl}) due to the first equation of system (\ref{phE}) and $f = f(\xi-\eta)$. 

If $f$ and $(p_1,q_1)$ are $L$-periodic functions in $x := \xi-\eta$ 
with period $L = 2 k K(k)$, the function $\phi_R$ and $(\hat{p}_1,\hat{q}_1)$ are non-periodic. When the second, linearly independent solution $(p,q) = (\hat{p}_1,\hat{q}_1)$ 
is used in the one-fold Darboux transformation (\ref{ODT}), it generates 
a new solution with an algebraic structure on the background of the rotational 
waves. The new solution approaches the rotational wave along the directions 
in the $(\xi,\eta)$ plane where $|\phi_R|$ grows to infinity. 

We recall (\ref{rot-wave}) and (\ref{rottrans}) to rewrite (\ref{growthrotexpl}) 
in the equivalent form:
\begin{equation}
\phi_R(\xi,\eta) = C + \frac{1}{2} (\xi + \eta) - \frac{H_0 k^3}{2 (1-k^2)}
\int_{0}^{k^{-1}(\xi-\eta)} {\rm dn}^2(z+K(k);k) dz.
\label{growthrotexp2}
\end{equation}
Also recall the complete elliptic integral of the second kind 
$$
E(k) = \int_0^{K(k)} {\rm dn}^2(z;k) dk.
$$
Over the period $L = 2k K(k)$, the integral in (\ref{growthrotexp2}) 
is incremented by $2 E(k)$, hence $|\phi_R(\xi,\eta)| \to \infty$ 
along every direction in the $(\xi,\eta)$-plane with the exception of the direction of the straight line:
\begin{equation}
\label{Omega}
\Omega := \left\{ (\xi,\eta) \in \mathbb{R}^2 \ \ : \quad 
(\xi + \eta) - \frac{H_0 k^2 E(k)}{(1-k^2) K(k)} (\xi - \eta) 
= 0 \right\}.
\end{equation}
The integration constant $C$ serves as a parameter which translates the straight line $\Omega$ in the $(\xi,\eta)$-plane within the period of the rotational wave.

Let us now take the one-fold Darboux transformation (\ref{ODT}) 
with the second linearly independent solution (\ref{hatform})  for the admissible eigenvalues $\lambda_1$ given by (\ref{Rdots}). By using the relations (\ref{squareeigen}), (\ref{hammmy-F}), and (\ref{useful123}), we obtain 
\begin{align}
\hat{w} &= w + \frac{4\lambda_1\hat{p}_1\hat{q}_1}{\hat{p}_1^2 + 
	\hat{q}_1^2}\nonumber\\
&= w + \frac{4\lambda_1 [p_1q_1 (\phi_R^2 w^2 -1) 
	+ \phi_R w (p_1^2 - q_1^2)]}{(p_1^2 + q_1^2)(\phi_R^2w^2 + 1)} \nonumber\\
&=  w + \frac{(4 H_0-w^2)(\phi_R^2 w^2 - 1) + 4 \phi_R w \partial_{\xi} w}{w (\phi_R^2 w^2 + 1)} \label{rogue1}
\end{align}
where $\hat{w} = -\hat{u}_{\xi}$ and $w = -u_{\xi}$. 

We show next that the new solution (\ref{rogue1}) describes a kink propagating on the background of the rotational wave. Indeed, the function $\phi_R(\xi,\eta) : \mathbb{R}^2 \mapsto \mathbb{R}$ 
is bounded and periodic in the direction of the line  
$\Omega$ given by (\ref{Omega}). In every other direction 
on the $(\xi,\eta)$-plane, $|\phi_R(\xi,\eta)| \to \infty$ so that the new solution  (\ref{rogue1}) satisfies the limit:
\begin{align} 
\lim_{ |\phi_R| \rightarrow \infty } \hat{w} 
=  w + \frac{4 H_0-w^2}{w} = \frac{4 H_0}{w},
\end{align} 
which coincides with (\ref{simpleODT}). As follows from (\ref{rottrans}), this limit is a half-period translated and reflected version 
of the rotational wave. Since the divergence of  $|\phi_R(\xi,\eta)| \to \infty$ is linear in $(\xi,\eta)$ as follows from (\ref{growthrotexp2}), 
the new solution (\ref{rogue1}) approaches the translated and reflected rotational wave algebraically fast.

\begin{figure}[htb!]
	\includegraphics[width=0.5\textwidth]{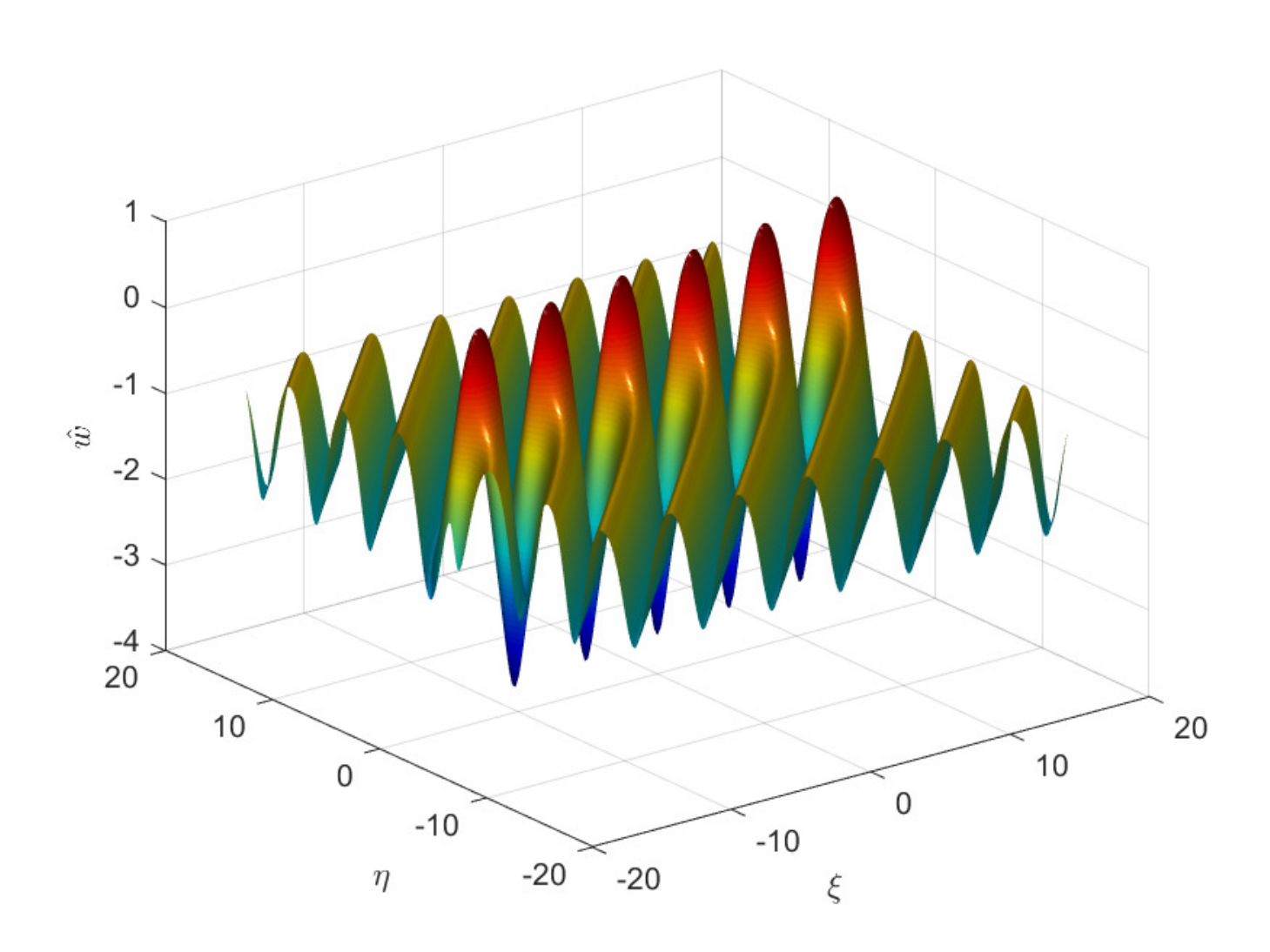}
	\includegraphics[width=0.45\textwidth]{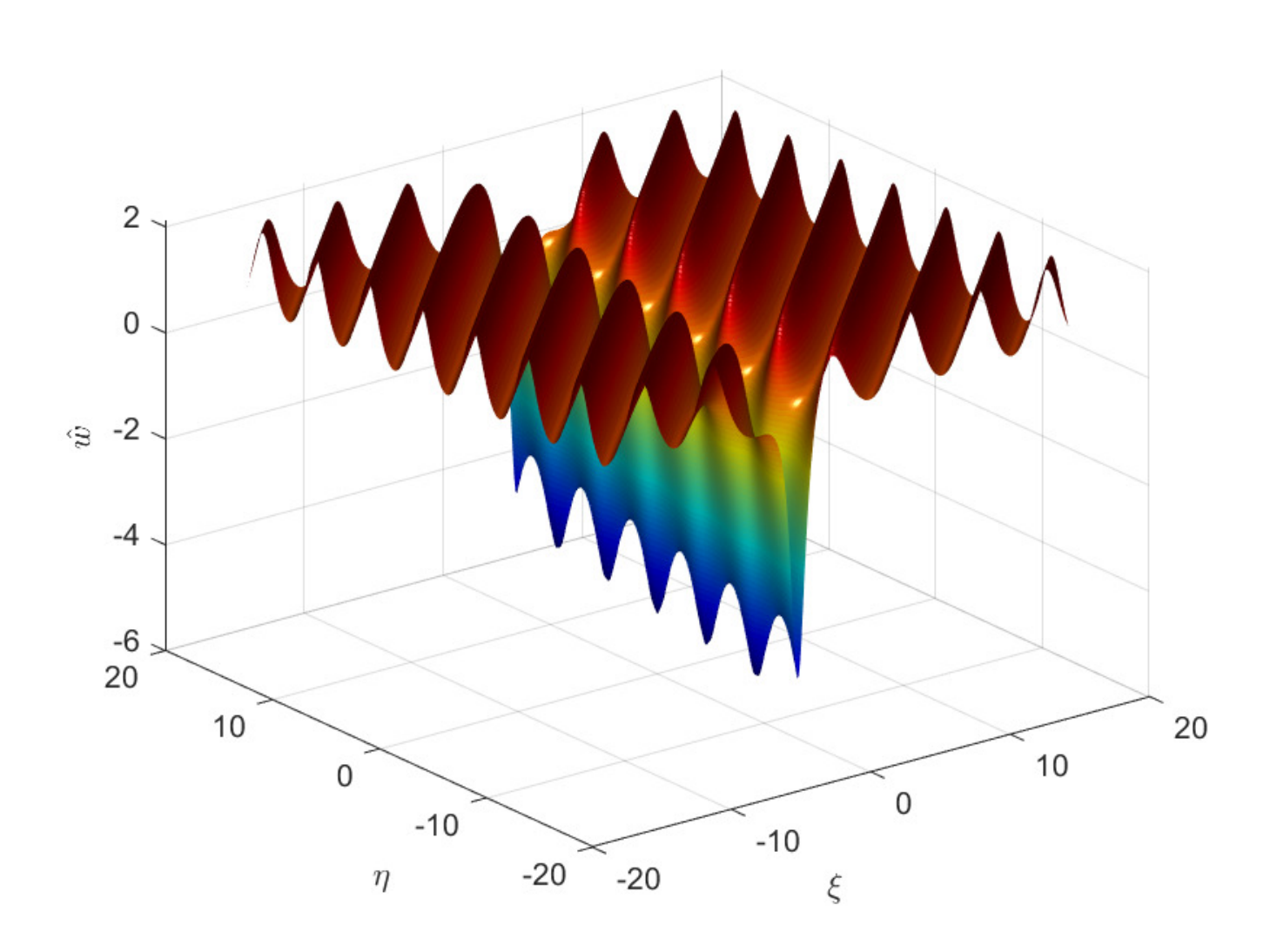}
	\caption{Localized waves on the rotational wave with $k = 0.95$
		generated from the one-fold Darboux transformation using eigenvalues
		(\ref{Rdots}) with the lower (left) and upper (right) signs.}\label{solits} 
\end{figure} 

Along the direction $\Omega$, the new solution (\ref{rogue1}) 
does not approach the rotational wave. It follows from (\ref{rogue1}) at the critical point of $w = -f'$, where $\partial_{\xi} w = -f''$ is zero, that   the maximum of $|\hat{w}|$ happens at the points, where $\phi_R = 0$ and
\begin{align}
\hat{w}  |_{\phi_R = 0} = w - \frac{4H_0-w^2}{w} = 
2 w - \frac{4H_0}{w}.   \label{maxsol} 
\end{align}  
Compared to the maximum of the rotational wave $\sup_{(\xi,\eta)\in \mathbb{R}^2} |w(\xi,\eta)| = 2 k^{-1}$, the maximum of the new 
solution  (\ref{rogue1}) is attained at $\sup_{(\xi,\eta)\in \mathbb{R}^2} |\hat{w}(\xi,\eta)| = 2 k^{-1} M$, where $M$ is the magnification factor given by 
\begin{align}
M(k) =  2 \mp \sqrt{1-k^2}.  \label{solmagnification}
\end{align}
The sign choice in (\ref{solmagnification}) corresponds to the sign choice in (\ref{Rdots}) and (\ref{Rdots-F}). The magnification factor $M$ determines the maximum of the localized wave propagating on the background of the rotational waves in the direction of the straight line $\Omega$. Position of the localized wave is changed by the parameter $C$ 
for the integration constant. The localized wave is greater for 
the lower sign in (\ref{Rdots}) and (\ref{Rdots-F}). 
Note that the magnification factor 
in (\ref{solmagnification}) was previously derived 
for similar solutions to the NLS and mKdV equations in \cite{CPnls,CPkdv}.

Figure \ref{solits} illustrates the exact solution (\ref{rogue1}) for $k = 0.95$ and two sign choices in (\ref{Rdots}). The value of $C$ is set to $0$ in (\ref{growthrotexp2}). We see numerically that the solution surface $|\hat{w}(\xi,\eta)|$ achieves its maximum at $(\xi,\eta)=(0,0)$ and is repeated along the direction of $\Omega$. This is the direction 
of propagation of the localized wave on the background of the rotational waves. The localized wave has a bigger magnification for the larger value of $\lambda_1$ (right panel) and smaller magnification for the smaller value of $\lambda_1$ (left panel).

By using the same solution formula (\ref{rogue1}), 
we have computed $\sin(\hat{u}) = \hat{u}_{\xi \eta}$ by numerically 
differentiating $\hat{w} = -\hat{u}_{\xi}$ in $\eta$ with a forward difference. The corresponding surface plots of $\sin(\hat{u})$ in $(x,t)$ are presented on Figure \ref{sinRotational}. Note that the kink and antikink propagate into opposite directions for the different sign choices of $\lambda_1$ in (\ref{Rdots}). Indeed, it follows from (\ref{Rdots-F}) and (\ref{Omega}) in variables $(x,t)$ that the kink and antikink propagate along the straight lines 
\begin{equation}
\label{kink-antikink}
x = \pm \frac{E(k)}{\sqrt{1-k^2} K(k)} t,
\end{equation}
hence the propagation directions are opposite to each other. 
Since $E(k) > \sqrt{1-k^2} K(k)$, the speed of propagation exceeds one, hence these solutions are relevant for the superluminal dynamics of the sine--Gordon equation (\ref{SG}).

\section{New solutions on the background of librational waves}

If the new solution $\hat{u} = \hat{u}(\xi,\eta)$ to the sine--Gordon equation (\ref{sine-gordon-en}) is given by the 
one-fold Darboux transformation (\ref{rogue1}) and $u = u(\xi,\eta)$ is the librational 
wave, then $\hat{u}$ is no longer real-valued because $H_0$ and $\lambda_1$ are complex-valued in (\ref{Ldots}) and (\ref{Ldots-F}).
The two-fold Darboux transformation is required to generate new real-valued
solutions on the background of the librational waves.

Let $(p_1,q_1)$ and $(p_2,q_2)$ be solutions to 
the linear equations (\ref{linearlaxnonlab}) and (\ref{linearlaxnonlab2}) with 
fixed values of $\lambda = \lambda_1$ and $\lambda = \lambda_2$ such that $\lambda_1 \neq \pm \lambda_2$. As is shown in \cite{CPkdv}, the two-fold Darboux transformation takes the form:
\begin{equation}\label{ODT2}
\hat{w} = w + \frac{4(\lambda_1^2 - 
	\lambda_2^2)[\lambda_1p_1q_1(p_2^2+q_2^2)-\lambda_2p_2q_2(p_1^2+q_1^2)]}
{(\lambda_1^2+\lambda_2^2)(p_1^2+q_1^2)(p_2^2+q_2^2)-2\lambda_1\lambda_2 
	[4p_1q_1p_2q_2+(p_1^2-q_1^2)(p_2^2-q_2^2)]} ,
\end{equation}
where $w := -u_{\xi}$ and $\hat{w} := -\hat{u}_{\xi}$. We take $\lambda_1$ and $H_0$ as in (\ref{Ldots}) and (\ref{Ldots-F}), and define $\lambda_2 = \bar{\lambda}_1$ 
with $p_2 = \bar{p}_1$ and $q_2 = \bar{q}_1$. By using 
(\ref{squareeigen}), (\ref{hammmy-F}), (\ref{useful123}), and (\ref{alg4}),
we obtain 
\begin{align}
\hat{w} = w + \frac{4 (\lambda_1^2 - \bar{\lambda}_1^2)(H_0-\bar{H}_0) w}
{(\lambda_1^2 + \bar{\lambda}_1^2) w^2 -2[-4 H_0^2 + \frac{1}{4} w^4 + (w')^2]}  = -w. \label{backgroundref12}  
\end{align}
The new solution (\ref{backgroundref12}) is simply a reflected version of the librational wave. 
Therefore, we are looking for the second, linearly independent solution 
to the linear equations (\ref{linearlaxnonlab}) and (\ref{linearlaxnonlab2}) for the same value of $\lambda_1$. One representation for the second 
solution is given by (\ref{hatform}). However, $w = p_1^2 + q_1^2 = -f'$ crosses zero for librational waves, hence the representation 
(\ref{hatform}) becomes singular at some points. For librational waves, 
we should define the second solutions in a different form 
used in \cite{CPkdv}:
\begin{equation}\label{eigfunc2}
\hat{p}_1 = \frac{\phi_L - 1}{q_1}, \quad 
\hat{q}_1 = \frac{\phi_L + 1}{p_1},
\end{equation}
where the function $\phi_L = \phi_L(\xi,\eta)$ 
satisfies the system of scalar equations:
\begin{equation}
\frac{\partial \phi_L}{\partial \xi} = 
\frac{f' (p_1^2-q_1^2)}{2p_1 q_1} \phi_L - 
\frac{f'(p_1^2+q_1^2)}{2p_1q_1}, \;\; 
\lambda_1 \frac{\partial \phi_L}{\partial \eta} = 
\frac{(p_1^2+q_1^2) \sin(f)}{2p_1q_1} \phi_L - \frac{(p_1^2-q_1^2) \sin(f)}{2p_1q_1}.
\label{phiEEE-prelim}
\end{equation} 
The representation (\ref{eigfunc2}) is non-singular because if either $p_1$ or $q_1$ vanish in some points, then equations (\ref{squareeigen}) and 
(\ref{useful123}) yield a contradiction with real $f$ and complex 
$\lambda_1$.  
As we prove below, the exact expression for $\phi_L$ is given by 
\begin{equation}
\phi_L(\xi,\eta) = (4 H_0 - (f')^2)\left(C + \frac{ \eta}{2\lambda_1} + 
\int_{0}^{\xi-\eta}\frac{2\lambda_1  
		(f')^2 dx}{(4 H_0 -(f')^2)^2}\right),\label{growthlib}
\end{equation}
where $C$ is an arbitary constant of integration.
By substituting  (\ref{squareeigen}), (\ref{hammmy-F}), and (\ref{useful123}) 
in (\ref{phiEEE-prelim}), we obtain:
\begin{equation}
\frac{\partial \phi_L}{\partial \xi} = 
\frac{2 f' f''}{(f')^2- 4H_0} \phi_L - 
\frac{2\lambda_1 (f')^2}{(f')^2- 4H_0}, \qquad 
\lambda_1 \frac{\partial \phi_L}{\partial \eta} = - 
\frac{2 \lambda_1 f' f''}{(f')^2- 4H_0} \phi_L + \frac{2(f'')^2}{(f')^2-4 H_0}.
\label{phiEEE}
\end{equation} 
By using 
\begin{equation}\label{covrot}
	\phi_L = (4 H_0 - (f')^2) \Upsilon
\end{equation}
with $\Upsilon = \Upsilon(\xi,\eta)$, system (\ref{phiEEE}) can be simplified to the form:
\begin{equation}\label{upsilon}
\frac{\partial \Upsilon}{\partial \xi} = \frac{2\lambda_1 (f')^2}{(4 H_0-(f')^2)^2}, \qquad 
\lambda_1 \frac{\partial \Upsilon}{\partial \eta} = 
-\frac{2 (f'')^2}{(4 H_0-(f')^2)^2}.
\end{equation} 
If follows from (\ref{alg4}) and (\ref{upsilon}) that 
\begin{equation}
\frac{\partial \Upsilon}{\partial \xi} + 
\frac{\partial \Upsilon}{\partial \eta}  = \frac{1}{2 \lambda_1} ,
\end{equation}
which implies that 
\begin{equation}\label{covrot2}
\Upsilon(\xi,\eta) = C + \frac{\eta}{2\lambda_1} + G(\xi-\eta)
\end{equation}
for some function $G(x) : \mathbb{R} \mapsto \mathbb{C}$ to be determined. Substituting this into (\ref{upsilon}) yields
	\begin{align*}
	G' = \frac{2\lambda_1 (f')^2}{(4 H_0-(f')^2)^2}, 
	\end{align*} 
so that integration and substitution into (\ref{covrot}) and (\ref{covrot2}) yields (\ref{growthlib}). 

The functions $f$ and $(p_1,q_1)$ are $L$-periodic functions with period 
$L = 4K(k)$ for librational waves, however, the functions $\phi_L$ 
and $(\hat{p}_1,\hat{q}_1)$ are non-periodic. We shall prove 
that $|\phi_L(\xi,\eta)| \to \infty$ as $|\xi| + |\eta| \to \infty$ 
everywhere in the $(\xi,\eta)$-plane. Indeed, by factoring out $\frac{1}{2\lambda_1}$ in the second term of equation 
	(\ref{growthlib}) and by using periodicity of $4 H_0 - (f')^2$, 
	we have $|\phi_L(\xi,\eta)| \to \infty$ if and only if 
	$|\tilde{\phi}_L(\xi,\eta)| \to \infty$, where 
	\begin{align*}
	\tilde{\phi}_L(\xi,\eta) &= \eta + 
\int_{0}^{\xi-\eta}\frac{4 \lambda_1^2 (f')^2}{((f')^2-4 H_0)^2} dx\\
	&= \eta + 
\int_{0}^{\xi-\eta}\frac{4 [(2k^2-1) + 2 i k \sqrt{1-k^2}] (f')^2}{((f')^2 + 4 ik\sqrt{1-k^2})^2} dx.
	\end{align*}
Taking the imaginary part yields 	
	\begin{align*}
	{\rm Im}[\tilde{\phi}] &= 8k\sqrt{1-k^2} \int_{0}^{\xi-\eta} 
	\frac{(f')^4 - 4 (2k^2-1) (f')^2 - 16 k^2(1-k^2)}{((f')^4+16 k^2(1-k^2))^2} (f')^2 dx \\
	&= 128 k^3 \sqrt{1-k^2} \int_{0}^{\xi-\eta} 
	\frac{k^2 {\rm cn}^4(x;k) + (1-2k^2) {\rm cn}^2(x;k) + k^2 -1 }{((f')^4+16 k^2(1-k^2))^2} (f')^2 dx \\
	&= -128 k^3 \sqrt{1-k^2} \int_{0}^{\xi-\eta} 
	\frac{{\rm sn}^2(x;k) {\rm dn}^2(x;k)}{((f')^4+16 k^2(1-k^2))^2} (f')^2 dx
	\end{align*}
	where we have used (\ref{libr-wave}) in order to express $f'(x) = 2 k {\rm cn}(x;k)$ and simplify the elliptic functions. 	
	The integrand  is clearly positive for every $k\in(0,1)$. This 
	means that ${\rm Im}[\tilde{\phi}]$ remains bounded 
	only in the diagonal direction on the $(\xi,\eta)$ plane, 
	however, in this direction ${\rm Re}[\tilde{\phi}]$ grows linearly in 
	$\eta$. Hence,   $|\phi_L(\xi,\eta)| \to \infty$ along every 
	direction in the $(\xi,\eta)$ plane.

Let us now take the two-fold Darboux transformation (\ref{ODT2}) with the second, linearly independent solution (\ref{eigfunc2}) to the linear equations (\ref{linearlaxnonlab}) and (\ref{linearlaxnonlab2}) 
for $\lambda_1 = k + i \sqrt{1-k^2}$ and $\lambda_2 = \bar{\lambda}_1$.
The new solution is written in the form:
\begin{equation}
\hat{w} = w + \frac{4(\lambda_1^2 - 
	\lambda_2^2)[\lambda_1\hat{p}_1\hat{q}_1(\hat{p}_2^2+\hat{q}_2^2)-\lambda_2\hat{p}_2\hat{q}_2
	(\hat{p}_1^2+\hat{q}_1^2)]}
{(\lambda_1^2+\lambda_2^2)(\hat{p}_1^2+\hat{q}_1^2)(\hat{p}_2^2+\hat{q}_2^2)-2\lambda_1\lambda_2
	[4\hat{p}_1\hat{q}_1\hat{p}_2\hat{q}_2+(\hat{p}_1^2-\hat{q}_1^2)(\hat{p}_2^2-\hat{q}_2^2)]}, \label{roguewave123}
\end{equation} 
where $(\hat{p}_2,\hat{q}_2)$ are taken as the complex conjugate to 
$(\hat{p}_1,\hat{q}_1)$.

\begin{figure}[htb!]
	\includegraphics[width=0.48\textwidth]{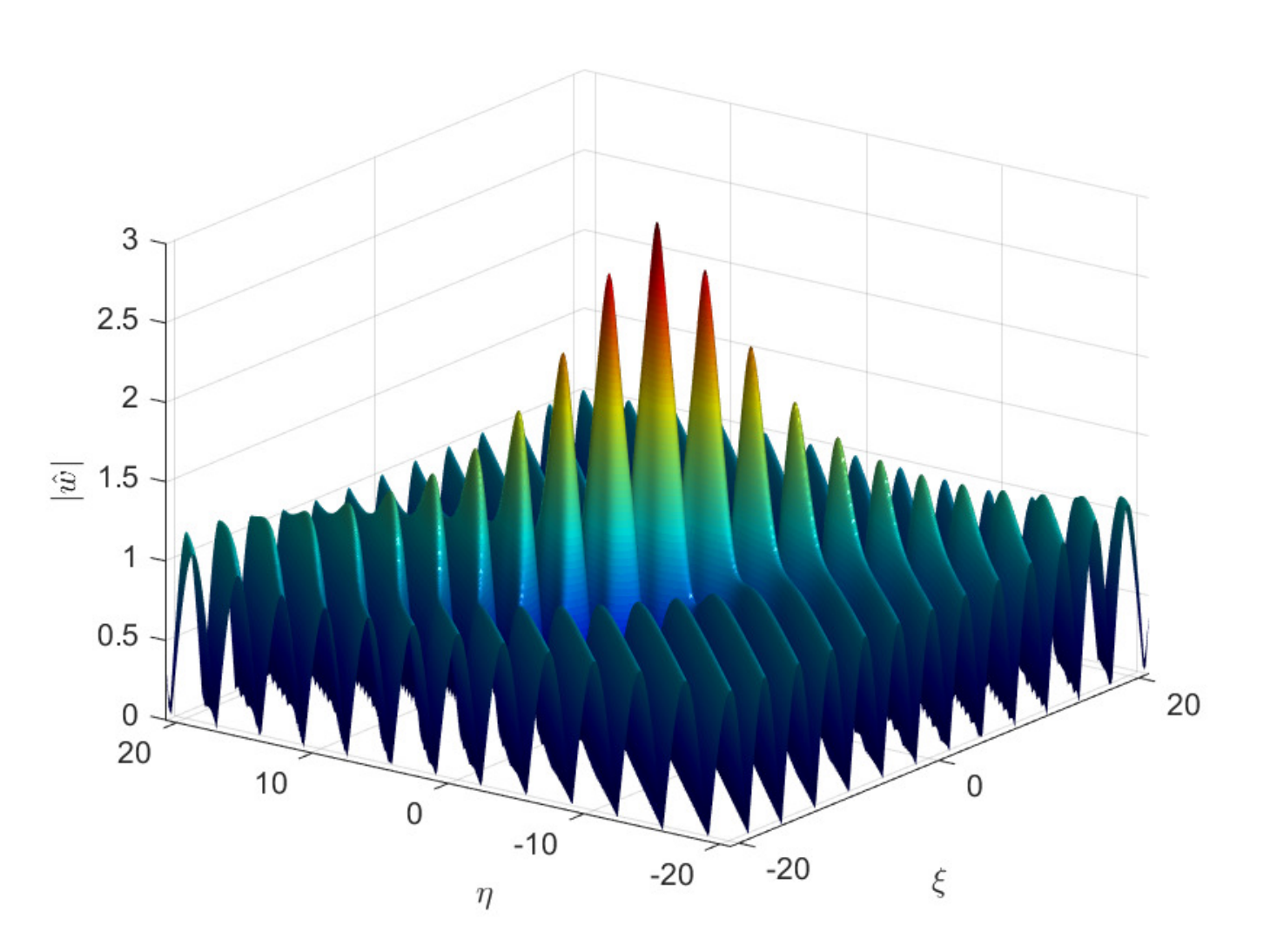}
	\includegraphics[width=0.48\textwidth]{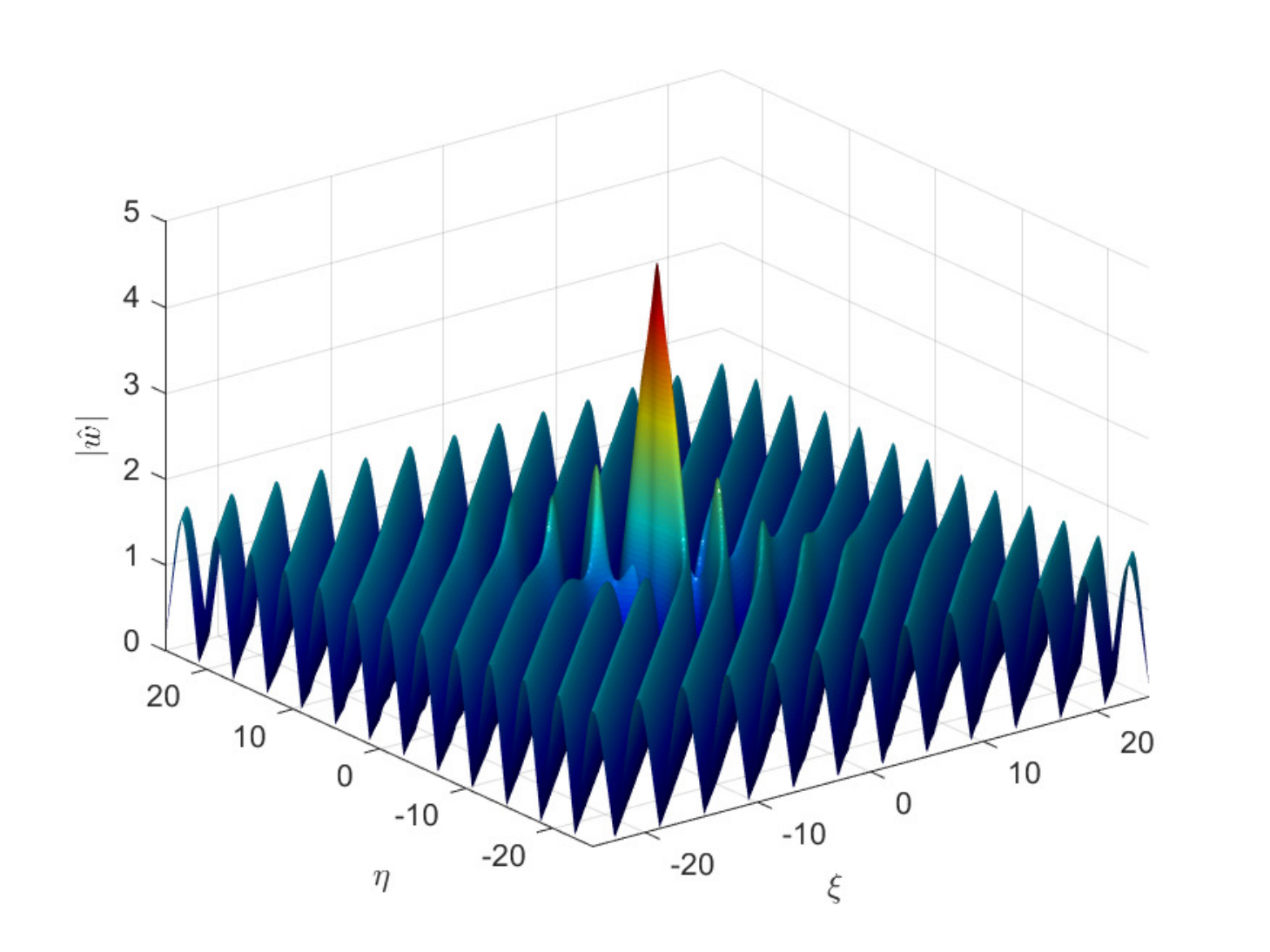}
	\caption{Rogue waves on the librational wave with $k=0.5$ (left) and 
		$k=0.8$ (right).} \label{rogueee}
\end{figure}

%\begin{figure}[htb!]
%	\includegraphics[width=0.48\textwidth]{lib_rogue_wave_k05}
%	\includegraphics[width=0.48\textwidth]{lib_rogue_wave_k08}
%	\caption{Rogue waves on the librational wave with $k=0.5$ (left) and 
%		$k=0.8$ (right).} \label{rogueee}
%\end{figure}

We will prove that the new solution (\ref{roguewave123}) describes an isolated rogue wave arising on the background of the librational wave. Indeed, the function $\phi_L(\xi,\eta) : \mathbb{R}^2 \mapsto \mathbb{R}$ 
is unbounded in every direction on the $(\xi,\eta)$ plane, so that 
\begin{align} 
\lim_{ |\phi_L| \rightarrow \infty } \hat{w} 
= w + \frac{4 (\lambda_1^2 - \bar{\lambda}_1^2)(H_0-\bar{H}_0) w}
{(\lambda_1^2 + \bar{\lambda}_1^2) w^2 -2[-4 H_0^2 + \frac{1}{4} w^4 + (w')^2]}  = -w,
\end{align} 
which coincides with (\ref{backgroundref12}). The divergence of  $|\phi_L(\xi,\eta)| \to \infty$ is again linear in $(\xi,\eta)$ as follows from (\ref{growthlib}), hence the new solution
(\ref{roguewave123}) approaches the reflected librational wave algebraically. 

It follows from (\ref{roguewave123}) at the critical points of $w = -f'$, where $\partial_{\xi} w = -f''$ is zero, that the maximum of $|\hat{w}|$ happens at the points, where $\phi_L = 0$ and
\begin{align}
\hat{w}  |_{\phi_L = 0} = w - \frac{4 (\lambda_1^2 - \bar{\lambda}_1^2)(H_0-\bar{H}_0) w}
{(\lambda_1^2 + \bar{\lambda}_1^2) w^2 -2[-4 H_0^2 + \frac{1}{4} w^4 + (w')^2]} = 3w.   \label{maxsol} 
\end{align}  
Compared to the maximum of the librational wave $\sup_{(\xi,\eta)\in \mathbb{R}^2} |w(\xi,\eta)| = 2 k$, the maximum of the new 
solution (\ref{roguewave123}) is attained at $\sup_{(\xi,\eta)\in \mathbb{R}^2} |\hat{w}(\xi,\eta)| = 6k$, hence the rogue wave has triple magnification compared to the background wave. Note that the rogue wave (\ref{roguewave123})
and the triple magnification factor was previously obtained 
for the mKdV equation in \cite{CPkdv}.

Figure \ref{rogueee} illustrates the exact solution (\ref{roguewave123}) 
for two particular values of $k$. The value of $C$ is set to $0$ 
in (\ref{growthlib}). It is clear that the solution surface of  $|\hat{w}(\xi,\eta)|$ achieves its maximum at $(\xi,\eta) = (0,0)$ 
where $\phi_L$ vanishes. The modulus is shown for a better resolution of the oscillations of the librational wave background.

Based on the same solution formula (\ref{roguewave123}), 
we have computed $\sin(\hat{u}) = \hat{u}_{\xi \eta}$ by numerically 
differentiating $\hat{w} = -\hat{u}_{\xi}$ in $\eta$ with a forward difference. The corresponding surface plots of $\sin(\hat{u})$ in $(x,t)$ are presented on Figure \ref{sinmiller} for different values of $k$.

\begin{figure}[htb!]
	\includegraphics[width=0.55\textwidth]{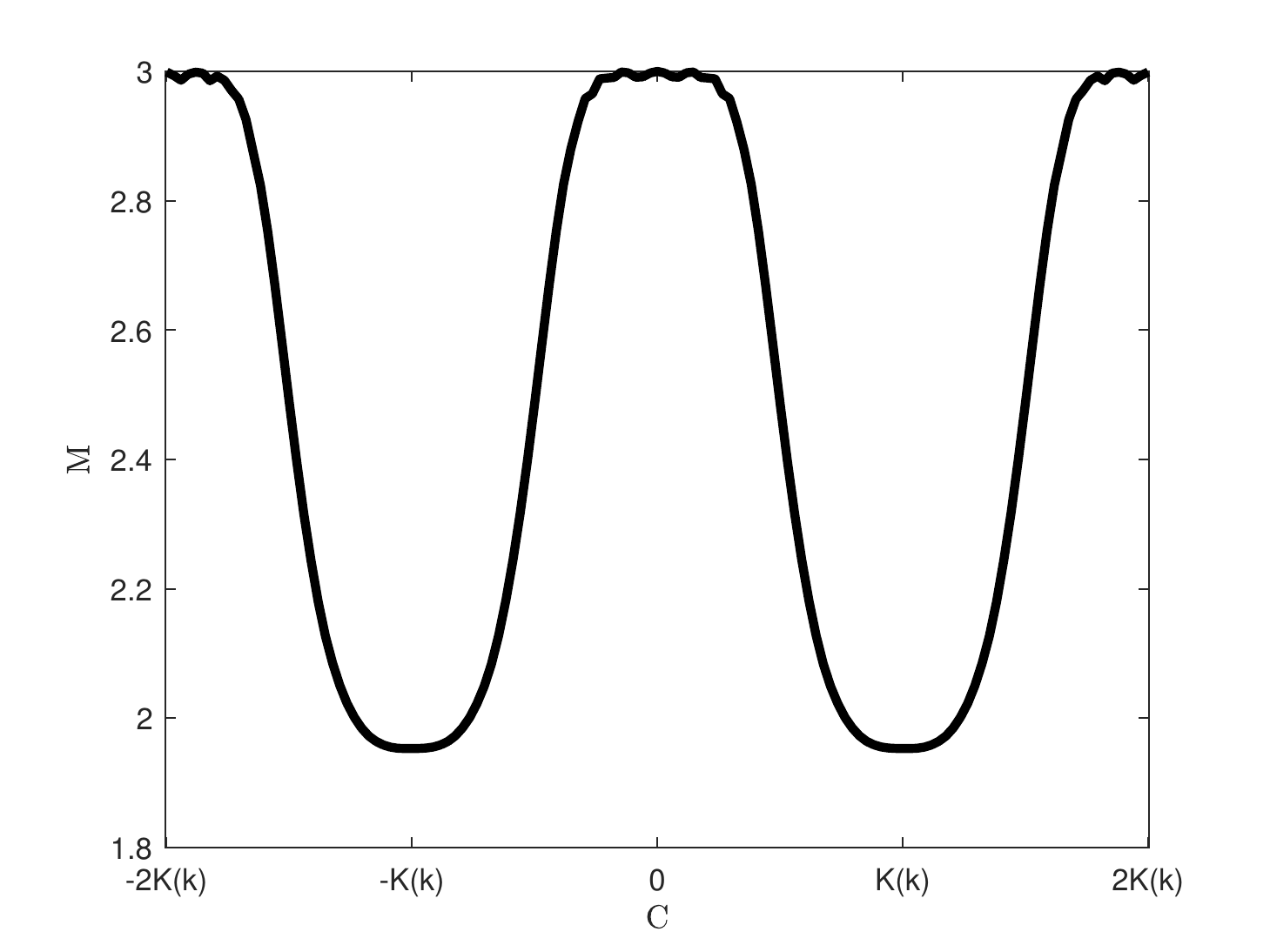}
	\caption{The magnification factor $M$ of the rogue wave $\hat{w}$ given by (\ref{roguewave123}) versus the constant of integration $C$ in (\ref{growthlib}) for $k=0.8$.}
	\label{constant_of_int}
\end{figure}

Finally, we inspect how the magnification of the rogue wave depends 
on the constant of integration $C$ in (\ref{growthlib}) and (\ref{roguewave123}). The magnification factor is defined as 
$$
M := \frac{\sup_{(\xi,\eta)\in \mathbb{R}^2} |\hat{w}(\xi,\eta)|}{\sup_{(\xi,\eta)\in \mathbb{R}^2} |w(\xi,\eta)|}.
$$
Figure \ref{constant_of_int} presents the plot of $M$
versus $C$ for $k = 0.8$. When $C = 0$, the magnification factor is maximal 
at $M = 3$. It is periodically continued with respect to $C$ 
and it reaches the minimal value below $2$. The minimal value of $M$ depends on $k$.

\section{Conclusion}

We have presented new solutions to the sine-Gordon equation using 
an algebraic method and the Darboux transformations. The new solutions 
describe localized structures on the background of rotational and librational waves. 
These localized structures are obtained for the particular eigenvalues 
of the linear Lax equations which correspond to bounded solutions 
in the space-time coordinates. The Darboux transformations 
use the second, linearly independent solutions to the linear Lax 
equations whcih are unbounded in space--time coordinates. 

For the rotational waves, the localized structure represents
a kink or an antikink propagating along a straight line. It appears 
from infinity and goes to infinity. This outcome is related to 
the modulational stability of the rotational waves. 

For the librational waves, the localized structure represents a 
rogue wave appearing from nowhere and disappearing without a trace. 
The rogue wave is related to the modulational instability of 
the librational waves. 

New solutions for localized structures on the background of rotational and librational waves can be used for modeling of dynamics of the fluxon condensates.
They represent the principal waveforms in the universal dynamics 
of the sine--Gordon equation arising in the semi-classical limit.

\vspace{0.25cm}

{\bf Acknowledgement.} The authors thank P.D. Miller and B.Y.Lu 
for sharing their preprint \cite{milluby} before submission 
and many relevant discussions. 
This project was supported in part by the National
Natural Science Foundation of China (No. 11971103).

\end{document}